\documentclass{aa}
\pdfoutput=1
\usepackage{hyperref}
\hypersetup{hidelinks}
\usepackage{natbib}
\bibpunct{(}{)}{;}{a}{}{,}
\usepackage{txfonts}

\usepackage{graphicx}
\usepackage{booktabs}

\usepackage[per-mode=reciprocal]{siunitx}
\DeclareSIUnit\solarmass{\ensuremath{M_{\odot}}}
\DeclareSIUnit\solarlum{\ensuremath{L_{\odot}}}
\DeclareSIUnit\year{yr}

\defcitealias{Robotham2015}{R15}
\newcommand{\RO}{\citetalias{Robotham2015}}
\newcommand{\Mary}{Mary et al., (in prep.)}

\newcommand{\Eq}[1]{Eq.~(\ref{#1})}

\newcommand{\Fig}[1]{Fig.~\ref{#1}}
\newcommand{\Sec}[1]{\S\ref{#1}}

\newcommand{\samplesize}{179}
\newcommand{\samplehudf}{147}
\newcommand{\samplehdfs}{32}
\newcommand{\sampleHalpha}{47}
\newcommand{\sampleHbeta}{132}
\newcommand{\samplezmin}{0.11}
\newcommand{\samplezmean}{0.53}
\newcommand{\samplezmax}{0.91}

\newcommand{\sampleDfourthousandcut}{12}
\newcommand{\sampleEQWcuthalpha}{7}
\newcommand{\sampleEQWcuthbeta}{5}

\newcommand{\sampleelartefacts}{three}
\newcommand{\samplenophotomtery}{four}

\newcommand{\sampleXraytot}{16}
\newcommand{\sampleXray}{11}
\newcommand{\sampleXrayHa}{5}

\newcommand{\sig}{\ensuremath{0.46^{+0.04}_{-0.03}}}
\newcommand{\sigtext}{$\sigma_{\text{intr}}=\sig$}
\newcommand{\slope}{\ensuremath{0.79^{+0.05}_{-0.05}}}

\newcommand{\zevol}{\ensuremath{2.78^{+0.78}_{-0.78}}}

\newcommand{\intercpt}{\ensuremath{-0.77^{+0.04}_{-0.04}}}

\newcommand{\covmatrix}{
  0.003 & -0.019 & -0.001 & 0.000 \\
  -0.019 & 0.620 & 0.011 & 0.000 \\
  -0.001 & 0.011 & 0.002 & -0.000 \\
  0.000 & 0.000 & -0.000 & 0.001}

\newcommand{\corrsig}{\ensuremath{0.44^{+0.05}_{-0.04}}}
\newcommand{\corrsigtext}{$\sigma_{\text{intr}}=\corrsig$}
\newcommand{\corrslope}{\ensuremath{0.83^{+0.07}_{-0.06}}}
\newcommand{\corrslopetext}{$a=\corrslope$}
\newcommand{\corrzevol}{\ensuremath{1.74^{+0.66}_{-0.68}}}
\newcommand{\corrzevoltext}{$c=\corrzevol$}
\newcommand{\corrintercpt}{\ensuremath{-0.83^{+0.05}_{-0.05}}}

\newcommand{\corrcovmatrix}{
  0.004 & -0.016 & -0.002 & 0.002 \\
  -0.016 & 0.459 & 0.010 & -0.003 \\
  -0.002 & 0.010 & 0.002 & -0.001 \\
  0.002 & -0.003 & -0.001 & 0.002}

\newcommand{\lowmsamplesize}{148}
\newcommand{\lowmsampleexcl}{31}
\newcommand{\lowmsig}{\ensuremath{0.49^{+0.04}_{-0.04}}}
\newcommand{\lowmslope}{\ensuremath{0.79^{+0.08}_{-0.07}}}
\newcommand{\lowmzevol}{\ensuremath{3.39^{+0.91}_{-0.90}}}
\newcommand{\lowmintercpt}{\ensuremath{-0.73^{+0.04}_{-0.04}}}

\newcommand{\corrlowmsig}{\ensuremath{0.47^{+0.06}_{-0.05}}}
\newcommand{\corrlowmslope}{\ensuremath{0.83^{+0.10}_{-0.09}}}
\newcommand{\corrlowmzevol}{\ensuremath{2.22^{+0.75}_{-0.76}}}
\newcommand{\corrlowmintercpt}{\ensuremath{-0.79^{+0.05}_{-0.05}}}

\newcommand{\twodslope}{\ensuremath{0.89^{+0.05}_{-0.05}}}
\newcommand{\twodintercpt}{\ensuremath{-0.82^{+0.04}_{-0.04}}}
\newcommand{\twodsig}{\ensuremath{0.49^{+0.04}_{-0.04}}}

\newcommand{\lowzsample}{72}
\newcommand{\lowzslope}{\ensuremath{0.86^{+0.09}_{-0.08}}}
\newcommand{\lowzintercpt}{\ensuremath{-0.92^{+0.07}_{-0.07}}}
\newcommand{\lowzsig}{\ensuremath{0.57^{+0.07}_{-0.06}}}

\newcommand{\highzsample}{107}
\newcommand{\highzslope}{\ensuremath{0.84^{+0.07}_{-0.06}}}
\newcommand{\highzintercpt}{\ensuremath{-0.73^{+0.06}_{-0.06}}}
\newcommand{\highzsig}{\ensuremath{0.46^{+0.05}_{-0.05}}}

\newcommand{\simgridsize}{1260}
\newcommand{\simsamplesize}{100}
\newcommand{\simnomcruns}{30}

\newcommand{\Amatrixinv}{
  1.336 & 0.014 & -0.150 & 0.171 \\
  0.638 & 0.863 & 0.574 & -2.621 \\
  -0.178 & -0.008 & 1.175 & -0.185 \\
  0.285 & 0.009 & -0.044 & 1.091}
\newcommand{\tvector}{
  0.293 \\
  0.061 \\
  -0.194 \\
  0.236}

\newcommand{\rexampleslopetrue}{0.8}
\newcommand{\rexamplezevoltrue}{2.0}

\newcommand{\rexamplesigtrue}{0.5}

\newcommand{\recsamplesizenocut}{\simsamplesize}

\newcommand{\Halpha}{\ensuremath{\mathrm{H}\alpha\ \lambda 6563}}
\newcommand{\Hbeta}{\ensuremath{\mathrm{H}\beta\ \lambda 4861}}
\newcommand{\Hgamma}{\ensuremath{\mathrm{H}\gamma\ \lambda 4340}}

\newcommand{\Oii}{\ensuremath{\mathrm{[O\,\textsc{ii}]}\ \lambda 3727}}
\newcommand{\Oiiia}{\ensuremath{\mathrm{[O\,\textsc{iii}]}\ \lambda 4959}}

\newcommand{\Oiii}{\ensuremath{\mathrm{[O\,\textsc{iii}]}\ \lambda 4959,5007}}

\newcommand{\Nii}{\ensuremath{\mathrm{[N\,\textsc{ii}]}\ \lambda 6584}}

\newcommand{\Ha}{\ensuremath{\mathrm{H}\alpha}}
\newcommand{\Hb}{\ensuremath{\mathrm{H}\beta}}
\newcommand{\Hg}{\ensuremath{\mathrm{H}\gamma}}

\newcommand{\HII}{\textnormal{H\,\textsc{ii}}}

\newcommand{\OII}{\textnormal{[O\,\textsc{ii}]}}
\newcommand{\OIII}{\textnormal{[O\,\textsc{iii}]}}

\begin{document}

\title{The MUSE Hubble Ultra Deep Field Survey}

\subtitle{XI. Constraining the low-mass end of the stellar mass - star formation
  rate relation at $z<1$\thanks{Based on observations made with ESO telescopes
    at the La Silla Paranal Observatory under programme IDs ID 060.A-9100(C),
    094.A-2089(B), 095.A-0010(A), 096.A-0045(A), and 096.A-0045(B).}}

\author{Leindert~A.~Boogaard \inst{1}
  \and Jarle~Brinchmann \inst{1,2}
  \and Nicolas~Bouch\'e \inst{3}
  \and Mieke~Paalvast \inst{1}
  \and Roland~Bacon \inst{4}
  \and Rychard~J.~Bouwens \inst{1}
  \and Thierry~Contini \inst{3}
  \and Madusha~L.P.~Gunawardhana \inst{1}
  \and Hanae~Inami \inst{4}
  \and Raffaella~A.~Marino \inst{5}
  \and Michael~V.~Maseda \inst{1}
  \and Peter~Mitchell \inst{6}
  \and Themiya~Nanayakkara \inst{1}
  \and Johan~Richard \inst{4}
  \and Joop~Schaye \inst{1}
  \and Corentin~Schreiber \inst{1}
  \and Sandro~Tacchella \inst{5,7}
  \and Lutz~Wisotzki \inst{6}
  \and Johannes~Zabl \inst{3}}

\institute{Leiden Observatory, Leiden University, P.O. Box 9513, 2300 RA,
  Leiden, The Netherlands
  \and Instituto de Astrof{\'\i}sica e Ci{\^e}ncias do Espa{\c c}o, Universidade
  do Porto, CAUP, Rua das Estrelas, PT4150-762 Porto, Portugal
  \and IRAP (Institut de Recherche en Astrophysique et Planétologie), Université
  de Toulouse, CNRS, UPS, Toulouse, France
  \and CRAL, Observatoire de Lyon, CNRS, Universit\'e Lyon 1, 9 Avenue
  Ch. Andr\'e, F-69561 Saint Genis Laval Cedex, France
  \and Department of Physics, ETH
  Z\"urich,Wolfgang--Pauli--Strasse\,27, 8093\,Z\"urich,
  Switzerland,
  \and Leibniz-Institut f\"ur Astrophysik Potsdam (AIP),
  An der Sternwarte 16, D-14482 Potsdam, Germany
  \and Harvard-Smithsonian Center for Astrophysics 60 Garden St.,
  Cambridge, MA 02138
  \\~\\ \email{boogaard@strw.leidenuniv.nl}}

\titlerunning{Constraining the low-mass end of the $M_{*}$-SFR relation at
  $z<1$}
\authorrunning{L.~A.~Boogaard~et.~al.}

\date{Accepted July 20, 2018}

\abstract{Star-forming galaxies have been found to follow a relatively tight
  relation between stellar mass ($M_{*}$) and star formation rate (SFR),
  dubbed the `star formation sequence'.  A turnover in the sequence has been
  observed, where galaxies with $M_{*} < \SI{e10}{\solarmass}$ follow a
  steeper relation than their higher mass counterparts, suggesting that the
  low-mass slope is (nearly) linear.  In this paper, we characterise the
  properties of the low-mass end of the star formation sequence between
  $7 \leq \log M_{*}[\si{\solarmass}] \leq 10.5$ at redshift
  $\samplezmin < z < \samplezmax$.  We use the deepest MUSE observations of
  the \emph{Hubble} Ultra Deep Field and the \emph{Hubble} Deep Field South to
  construct a sample of \samplesize\ star-forming galaxies with high
  signal-to-noise emission lines.  Dust-corrected SFRs are determined from
  \Hbeta\ and \Halpha.  We model the star formation sequence with a Gaussian
  distribution around a hyperplane between $\log M_{*}$, $\log \text{SFR}$, and
  $\log (1+z)$, to simultaneously constrain the slope, redshift evolution, and
  intrinsic scatter.  We find a sub-linear slope for the low-mass regime where
  $\log \text{SFR}[\si{\solarmass\per\year}] = \corrslope\log
  M_{*}[\si{\solarmass}] + \corrzevol \log (1+z)$, increasing with redshift.
  We recover an intrinsic scatter in the relation of \corrsigtext\ dex, larger
  than typically found at higher masses.  As both hydrodynamical simulations
  and (semi-)analytical models typically favour a steeper slope in the
  low-mass regime, our results provide new constraints on the feedback
  processes which operate preferentially in low-mass halos.}

\keywords{Galaxies: star formation -- Galaxies: ISM -- Galaxies:
  formation -- Galaxies: evolution -- Methods: statistical}

\maketitle

\section{Introduction}
\label{sec:introduction}
How galaxies grow is one of the fundamental questions in astronomy.  The picture
that has emerged is that a galaxy builds up its stellar mass mainly through star
formation, which is triggered by gas accretion from the cosmic web
\citep[e.g.][]{DekelA_09a,VandeVoort2012}, while mergers with other galaxies
play only a minor role \citep[except for massive systems;][]{Bundy2009}.

In the past decade, star-forming galaxies have been found to form a reasonably
tight quasi-linear relation between stellar mass ($M_{*}$) and star formation
rate (SFR) \citep{Brinchmann2004, Noeske2007a, Elbaz2007, Daddi2007, Salim2007}
over a wide range of masses and out to high redshifts \citep{Pannella2009,
  Santini2009, Oliver2010, PengY_10a, Rodighiero2010, Karim2011, Bouwens2012,
  Whitaker2012, StarkD_13a, Whitaker2014, Ilbert2015, Lee2015, Renzini2015,
  Schreiber2015, Shivaei2015, Salmon2015, Tasca2015, Gavazzi2015,
  Kurczynski2016, Tomczak2016, Santini2017,Bisigello2017}, which is often
referred to as the `main sequence of star-forming galaxies' or the `star
formation sequence'.  In contrast, galaxies that are undergoing a starburst or
have already quenched their star formation respectively lie above and below the
relation.  This main sequence is close to a similar scaling relation for halos
\citep{BirnboimY_07a,NeisteinE_08a, GenelS_08a, FakhouriO_08a,Correa2015a,
  Correa2015b} where the growth rate increases super-linearly~\footnote{There is
  a tension between the shallow slope of the observed main sequence with the
  super-linear slope expected in models, which is set by the index of the
  initial dark matter power spectrum~\citep{BirnboimY_07a,
    NeisteinE_08a,Correa2015a,Correa2015b}.} with halo mass, and this has been
interpreted as supporting the picture where galaxy growth is driven by gas
accretion from the cosmic web \citep[e.g.][]{BoucheN_10a, Lilly2013,
  Rodriguez-PueblaA_16a, Tacchella2016a}.

This interpretation is supported by hydrodynamical simulations of galaxy
formation \citep{Schaye2010, Haas2013a, Haas2013b, Torrey2014, Hopkins2014,
  Crain2015, Hopkins2016}, where a global equilibrium relation is found between
the inflow and outflow of gas and star formation in galaxies.  In this picture
the star formation acts as a self-regulating process, where the inflow of gas,
through cooling and accretion, is balanced by the feedback from massive stars
and black holes \citep[e.g.][]{Schaye2010}.  Furthermore, semi-analytical
models \citep[e.g.][]{DuttonA_10a,Mitchell2014,Cattaneo2011,Cattaneo2017} and
relatively simple analytic theoretical models which connect the gas supply (from
the cosmological accretion) to the gas consumption can also reproduce the main
features of the main sequence rather well \citep[e.g.][]{BoucheN_10a,
  Dave2012, Lilly2013, DekelA_13a, DekelA_14a, MitraS_15a,
  Rodriguez-PueblaA_16a, Rodriguez-PueblaA_17a}~\footnote{For an alternative
  interpretation, cf. \cite{Gladders2013, Kelson2014, Abramson2016}.}.

The parameters of the $M_{*}$-SFR relation (i.e. slope, normalisation, and
scatter) are thus important, as they provide us with insight into the relative
contributions of different processes operating at different mass scales, in
particular when comparing the values of the parameters to their counterparts in
dark matter halo scaling relations.  The normalisation of the star formation
sequence is governed by the change in cosmological gas accretion rates and gas
depletion timescales.  The slope can be sensitive to the effect of various
feedback processes acting on the accreted gas, which prevent (or enhance) star
formation.  The intrinsic scatter around the equilibrium relation is
predominantly determined by the stochasticity in the gas accretion process
\citep[e.g.][]{ForbesJ_14a, MitraS_17a}, but can also be driven by dynamical
processes that rearrange the gas inside galaxies \citep{Tacchella2016a}.  The
$M_{*}$-SFR relation is observed to be reasonably tight, with an intrinsic
scatter of only $\approx 0.3$ dex \citep[][though we caution against a blind
comparison as different observables probe star formation on different
timescales]{Noeske2007a, Salmi2012, Whitaker2012, Guo2013,
  Speagle2014,Schreiber2015,Kurczynski2016}.  Yet, it has proven to be
challenging to place firm constraints on the intrinsic scatter as one needs to
deconvolve the scatter due to measurement uncertainty
\citep[e.g.][]{Speagle2014, Kurczynski2016, Santini2017}.

Observationally, the slope has been difficult to measure, particularly at the
low-mass end, as most studies have been sensitive to galaxies with stellar
masses above $\log M_*[\si{\solarmass}]\sim 10$ and often lack dynamical range
in mass.  In addition, while it is well known that there is significant
evolution in the normalisation of the sequence with redshift, most studies have
measured the slope in bins of redshift.  For a flux limited sample this could
introduce a bias in the slope because overlapping populations at different
normalisations are not sampled equally in mass within a single redshift bin.
The slope may also be mass dependent and indeed recent studies have observed
that the relation turns over around a mass of $M_{*} \sim \SI{e10}{\solarmass}$
\citep{ Whitaker2012, Whitaker2014, Lee2015, Schreiber2015, Tomczak2016} and
shows a steeper slope below the turnover mass.  In the low-mass regime, a
(nearly) linear slope has generally been expected
\citep[e.g.][]{Schreiber2015, Tomczak2016}, motivated also by the fact that
there is very little evolution in the faint-end slope of the blue stellar mass
function with redshift \citep{Peng2014}.  \cite{Leja2015} showed that the
sequence cannot have a slope $a < 0.9$ at all masses because this would lead
to a too high number density between $10 < \log M_{*}[\si{\solarmass}] < 11$ at
$z = 1$.

In addition to the observational challenges, careful modelling is required to get
reliable constraints on the parameters (slope, normalisation, scatter) of the
star formation sequence.  It is important to properly take selection effects
into account as well as the uncertainties on both the stellar masses and star
formation rates (and, if spectroscopy is lacking, also on the photometric
redshifts).  The latter in particular, due to the fact that there is intrinsic
scatter in the relation that needs to be deconvolved from the measurement
errors.  Common statistical techniques do not take these complications into
account self-consistently, which leads to biases in the results.

Putting the existing observations in perspective, it is clear that a large
dynamical range in mass is necessary to measure the slope of the star formation sequence
in the low-mass regime.  Deep field studies, that can blindly detect large
numbers of galaxies down to masses much below $\SI{e10}{\solarmass}$, are
invaluable in this regard \citep[e.g.][]{Kurczynski2016}.  Yet, such studies
are challenged by having to measure all observables, distances as well as
stellar masses and star formation rates, from the same photometry.  This can
lead to undesirable correlations between different observables.  At the same
time the measurements suffer from the uncertainties associated with photometric
redshifts.  Spectroscopic follow up is crucial in this regard, but can suffer
from biases due to photometric preselection.

With the advent of the \emph{Multi Unit Spectroscopic Explorer} (MUSE;
\citealt{Bacon2010}) on the VLT it is now possible to address these concerns.
With the deep MUSE data obtained over the \emph{Hubble} Ultra Deep Field
\citep[HUDF;][]{Bacon2017} and \emph{Hubble} Deep Field South
\citep[HDFS;][]{Bacon2015}, we can `blindly' detect star-forming galaxies in
emission lines down to very low levels ($\sim \SI{e-3}{\solarmass \per \year}$)
and obtain a precise spectroscopic redshift estimate at the same time
\citep{Inami2017}.  These data provides a unique view into the low-mass regime
of the star formation sequence.

In this paper we present a characterisation of the low-mass end of the
$M_{*}$-SFR relation, using deep MUSE observations of the HUDF and HDFS.  We
characterise the properties of the $M_{*}$-SFR relation down stellar masses of
$M_{*} \sim \SI{e8}{\solarmass}$ and SFR $\sim \SI{e-3}{\solarmass \per \year}$,
out to $z < 1$, and trace the SFR in individual galaxies with masses as low as
$M_{*} \la \SI{e7}{\solarmass}$ at $z\sim0.2$.  We model the relation using a
self-consistent Bayesian framework and describe it with a Gaussian distribution
around a plane in (log mass, log SFR, log redshift)-space.  This allows us to
simultaneously constrain the slope and evolution of the star formation sequence
as well as the amount of intrinsic scatter, while taking into account
heteroscedastic errors (i.e. a different uncertainty for each data point).

The structure of the paper is as follows.  In \Sec{sec:observations-methods} we
first introduce the MUSE data set and outline the selection criteria used to
construct our sample of star-forming galaxies.  We then go into the methods used
to determine a robust stellar mass and a SFR from the observed emission lines.
Before looking at the results, we discuss the consistency of our SFRs in
\Sec{sec:cons-sfr-indic}.  We then introduce the framework of our Bayesian
analysis used to characterise the $M_{*}$-SFR relation (\Sec{sec:modelling}) and
present the results in \Sec{sec:results}.  We discuss the robustness of the
derived parameters in \Sec{sec:selection-function-completeness}.  Finally, we
discuss our results in the context of the literature and models, and the
physical implications (\Sec{sec:discussion}).  We summarise with our conclusions
in \Sec{sec:conclusions}.  Throughout this paper we assume a \cite{Chabrier2003}
stellar initial mass function and a flat $\Lambda$CDM cosmology with
$H_0 = \SI{70}{km.s^{-1}.Mpc^{-1}}$, $\Omega_m = 0.3$ and
$\Omega_{\Lambda} = 0.7$.

\begin{figure}
  \centering
  \includegraphics[width=\columnwidth]{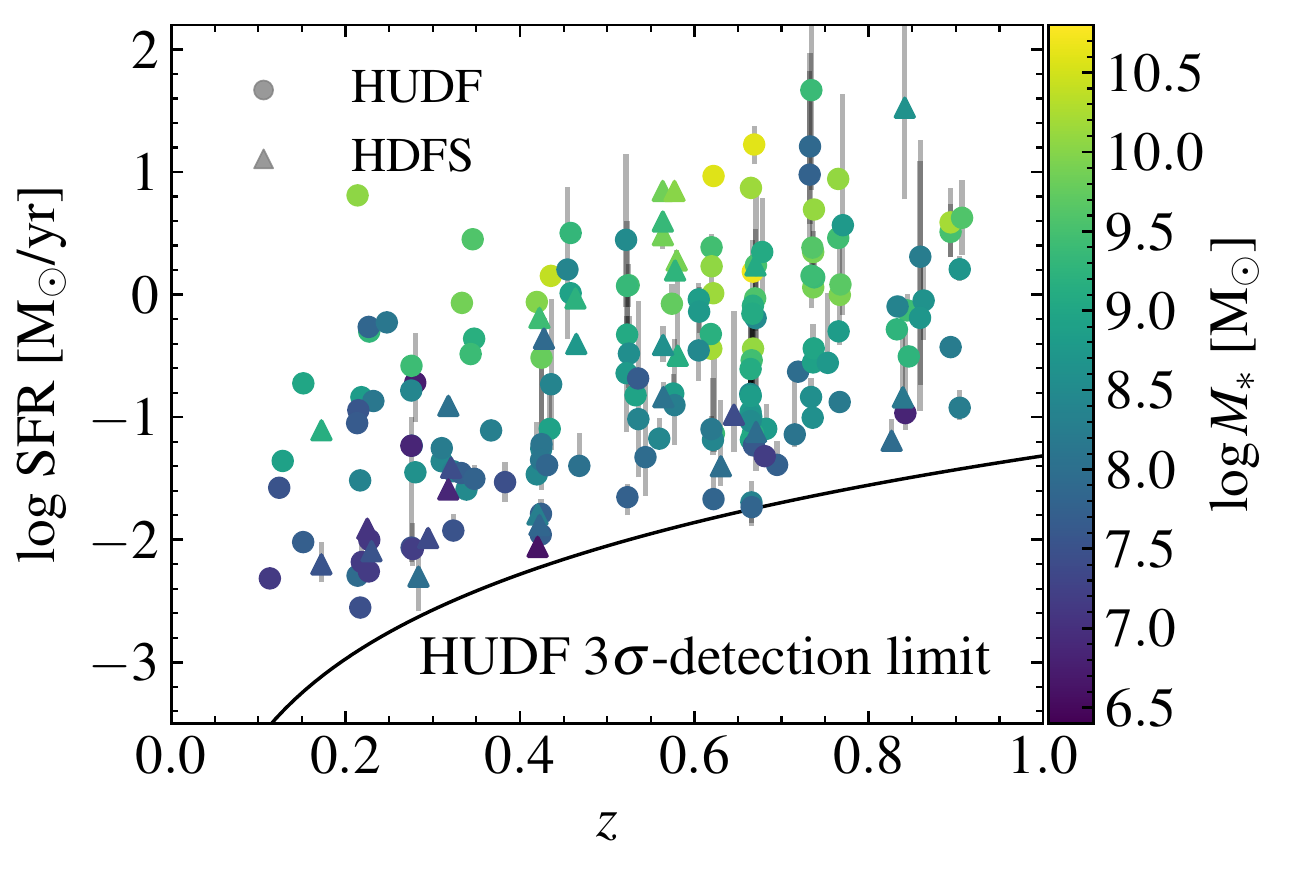}
  \caption{\label{fig:redshift-sfct} Redshift distribution of our galaxies
    plotted against their (dust-corrected) SFR (1$\sigma$ error bars are in
    grey).  The colour denotes the stellar mass.  The solid line depicts the
    lowest uncorrected SFR from \Hbeta\ we can detect in the HUDF at each
    redshift (which is effectively determined by the requirement that
    S/N$(\Hgamma) > 3$; see \protect\Sec{sec:star-formation-rates}).}
\end{figure}

\section{Observations and methods}
\label{sec:observations-methods}
\label{sec:data}
To study the properties of the galaxy population down to low masses and star
formation rates, deep spectroscopic observations are required for a large number
of sources.  We exploit the unique observations taken with the MUSE instrument
over the \emph{Hubble} Ultra Deep Field \citep{Bacon2017} and the \emph{Hubble}
Deep Field South \citep{Bacon2015} to investigate the star formation rates in
low-mass galaxies at $\samplezmin < z < \samplezmax$.  We provide a brief
presentation of the observations and data reduction in the next section, but
refer to the corresponding papers for details.

The MUSE instrument is an integral-field spectrograph situated at UT4 of the
Very Large Telescope.  It has a field-of-view of
$\SI{1}{\arcmin} \times \SI{1}{\arcmin}$ when operating in wide-field-mode,
which is fed into 24 different integral-field units.  These sample the
field-of-view at \SI{0.2}{\arcsec} resolution.  The spectrograph covers the
spectrum across 4650\AA\ - 9300 \AA\ with a spectral resolution of
$R \equiv \lambda / \Delta \lambda \simeq 3000$.  The result of a MUSE
observation is a data cube of the observed field, with two spatial and one
spectral axes, i.e. an image with spectroscopic information at every pixel.

\subsection{Observations, data reduction, and spectral line fitting}
The HUDF \citep{Beckwith2006} was observed with MUSE in a layered strategy.  The
deepest region consists of a single $1'\times 1'$ pointing with a total
integration depth of 31 hours.  This deep region lies embedded in a larger
$3'\times 3'$ mosaic consisting of 9 individual MUSE pointings, each of which is
10 hours deep.  The average full width at half maximum (FWHM) seeing measured in
the data cubes is \SI{0.6}{\arcsec} at \SI{7750}{\angstrom}.  For the purpose of
this work we use all galaxies from the \textsf{mosaic} region, including the
deep (\textsf{udf10}) region, which we refer to collectively as the (MUSE) HUDF.

Because of its similar depth, we also include the MUSE observation of the HDFS
\citep{Williams2000} which was observed as part of the commissioning activities.
These observations consist of a single deep field ($1' \times 1'$) with a total
integration time of 27 hours and a median seeing of \SI{0.7}{\arcsec}.

The full data acquisition and reduction of the HUDF is detailed in
\cite{Bacon2017} (for a description of the MUSE data reduction pipeline see
Weilbacher et al., in prep.).  The data reduction of the HUDF is essentially
based on the reduction of the HDFS, which is detailed in \cite{Bacon2015}, with
several improvements.  We use HUDF version 0.42 and HDFS version 1.0, which
reach a $3\sigma$-emission line depth for a point source (\SI{1}{\arcsec}) of
1.5 and $3.1 \times 10^{-19}$ \si{erg.s^{-1}.cm^{-2}} (\textsf{udf10} and
\textsf{mosaic}) and \SI{1.8e-19}{erg.s^{-1}.cm^{-2}} (HDFS), measured between
the OH skylines at \SI{7000}{\angstrom}.

Sources in the HUDF were identified using both a blind and a targeted approach.
The latter uses the sources from the UVUDF catalogue \citep{Rafelski2015} as
prior information to extract objects from the MUSE cube.  A blind search of the
full cube was also conducted, using a tool specifically developed for MUSE cubes
called \textsf{ORIGIN} (\cite{Bacon2017}; \Mary).  A similar approach was
already followed for the HDFS.  Here sources were identified based on the
\cite{Casertano2000} catalogue and blind emission line searches of the data cube
were done with the automatic detection tools
\textsf{Muselet}~\footnote{\url{http://mpdaf.readthedocs.io/en/latest/muselet.html}}
and \textsf{LSDCat} \citep{Herenz2017} as well as through visual inspection, and
cross-correlated with the corresponding photometric catalogue, as described in
\cite{Bacon2015}.

The process of determining redshifts and constructing a full catalogue from the
extracted sources is described in \cite{Inami2017} for the HUDF (and a similar
approach was followed for the HDFS).  In short, redshifts were determined
semi-automatically by cross-matching template spectra with the identified
sources and subsequently inspected and confirmed by at least two independent
investigators.  For emission line galaxies an additional constraint comes from
the requirement that the emission line flux is coherent in a narrow band image
around the line in the MUSE cube.  The typical error on the MUSE spectroscopic
redshifts is $\sigma_{z} = 0.00012(1+z)$ \citep{Inami2017}, which we will take
into account in the modelling (conservatively taking
$\sigma_{\log(1+z)} = 0.0005$ for all galaxies; \Sec{sec:modelling})

For all detected sources one dimensional spectra are extracted using a straight
sum extraction over an aperture around each source (based on the MUSE point
spread function convolved with the \cite{Rafelski2015} segmentation map, see
\citealt{Bacon2017}).  From the extracted 1D spectra emission line fluxes are
fitted in velocity space, using an updated version of the \textsf{Platefit} code
described in \cite{Tremonti2004} and \cite{Brinchmann2004, Brinchmann2008}.
\textsf{Platefit} assumes a Gaussian line profile for all lines, with the same
intrinsic width and velocity.  The result is a measurement of the flux and
equivalent width of all emission lines present, with the uncertainties obtained
from propagating the original pipeline errors.  We define the signal-to-noise
(S/N) in a particular spectral line as the line flux over the line flux error.
We also determine the strength of the 4000 \AA\ break, $D_{n}(4000)$, measured
over $3850-3950$\AA\ and $4000-4100$\AA\ \citep{Kauffmann2003}.  We note that
the stellar absorption underlying the emission lines is taken into account by
\textsf{Platefit}.

\subsection{Sample selection}
\label{sec:sample-selection}
From the HUDF and HDFS catalogues we construct our sample of star-forming
galaxies using the following constraints:

\begin{enumerate}
\item We use \Hbeta\ or \Halpha\ to derive the SFR (see
  \Sec{sec:star-formation-rates}) and in either case we always need \Hbeta\ (to
  directly probe the SFR or to correct for dust extinction in \Halpha).  As a
  result, we are limited to the range of redshifts where \Hbeta\ falls within
  the MUSE spectral range.  Subsequently, we only take objects into account that
  have a redshift $z < (9300 / 4861) - 1 = 0.913$.
\item In order to derive a robust SFR and dust correction, we only allow objects
  with a signal-to-noise ratio $>3$ in the relevant pair of Balmer lines.  This
  means S/N $>3$ in either \Hbeta\ and \Hgamma\ (for \Hbeta\ derived SFRs) or
  \Halpha\ and \Hbeta\ (for \Halpha\ derived SFRs).
\end{enumerate}

Included in the above criteria are some galaxies that are not actively
star-forming and lie on the `red-sequence'.  Since these galaxies are not
expected to lie on the $M_{*}$-SFR relation, we exclude them from the analysis
based on their spectral features:
\begin{enumerate}
\setcounter{enumi}{2}
\item We remove \sampleDfourthousandcut\ galaxies with a strong 4000 \AA\ break
  by only allowing galaxies with a $D_{n}(4000) < 1.5$.
\item We omit galaxies with a rest-frame equivalent width in either \Halpha\ or
  \Hbeta\ of < 2\AA~\footnote{Following the convention that emission-line
    equivalent widths (EQW) are negative, this translates to excluding EQW >
    $-2$\AA.}.  This removed an additional \sampleEQWcuthalpha\ and
  \sampleEQWcuthbeta\ objects, respectively.
\end{enumerate}

In addition, \sampleelartefacts\ sources were removed from the sample due to
severe artefacts in their emission lines (see \Sec{sec:cons-sfr-indic}).  All
sources selected based on the MUSE data are detected in the \emph{HST} imaging.
However, \samplenophotomtery\ sources were removed because there photometry was
severely blended, prohibiting a mass estimate.

\begin{enumerate}
  \setcounter{enumi}{4}
\item We remove potential AGN from our sample in the HUDF by cross-matching our
  sources with the \emph{Chandra} Deep Field South 7Ms X-ray catalogue
  \citep{Luo2016}.  We also confirm the location of the sources in the
  star-forming region of different emission line diagnostic diagrams.
\end{enumerate}
A total of \sampleXraytot\ galaxies with $z<0.913$ from the MUSE catalogue are
detected in X-rays.  Five of these sources (including one AGN) show passive
spectra without emission lines and did not pass the previous criteria.
Cross-matching our star-forming sample (after applying criteria 1 through 4)
left \sampleXray\ galaxies that were detected in X-rays.  Five of these sources
(ID\#855, 861, 863, 895, and 902) are in the \Ha-subsample and six (ID\#867, 869,
874, 875, 884, and 905) are in the \Hb-subsample.  All of these sources were
classified as `Galaxy' in the \cite{Luo2016} catalogue (according to their 6
criteria based on X-ray luminosity, spectral index, flux-ratios and previous
spectroscopic identification), except for ID\# 875 which was classified as an
AGN and which we subsequently removed from the sample.  \cite{Luo2016} caution
however that sources classified as `Galaxy' may still host low-luminosity or
heavily obscured AGN.

\begin{figure}
  \centering
  \includegraphics[width=\columnwidth]{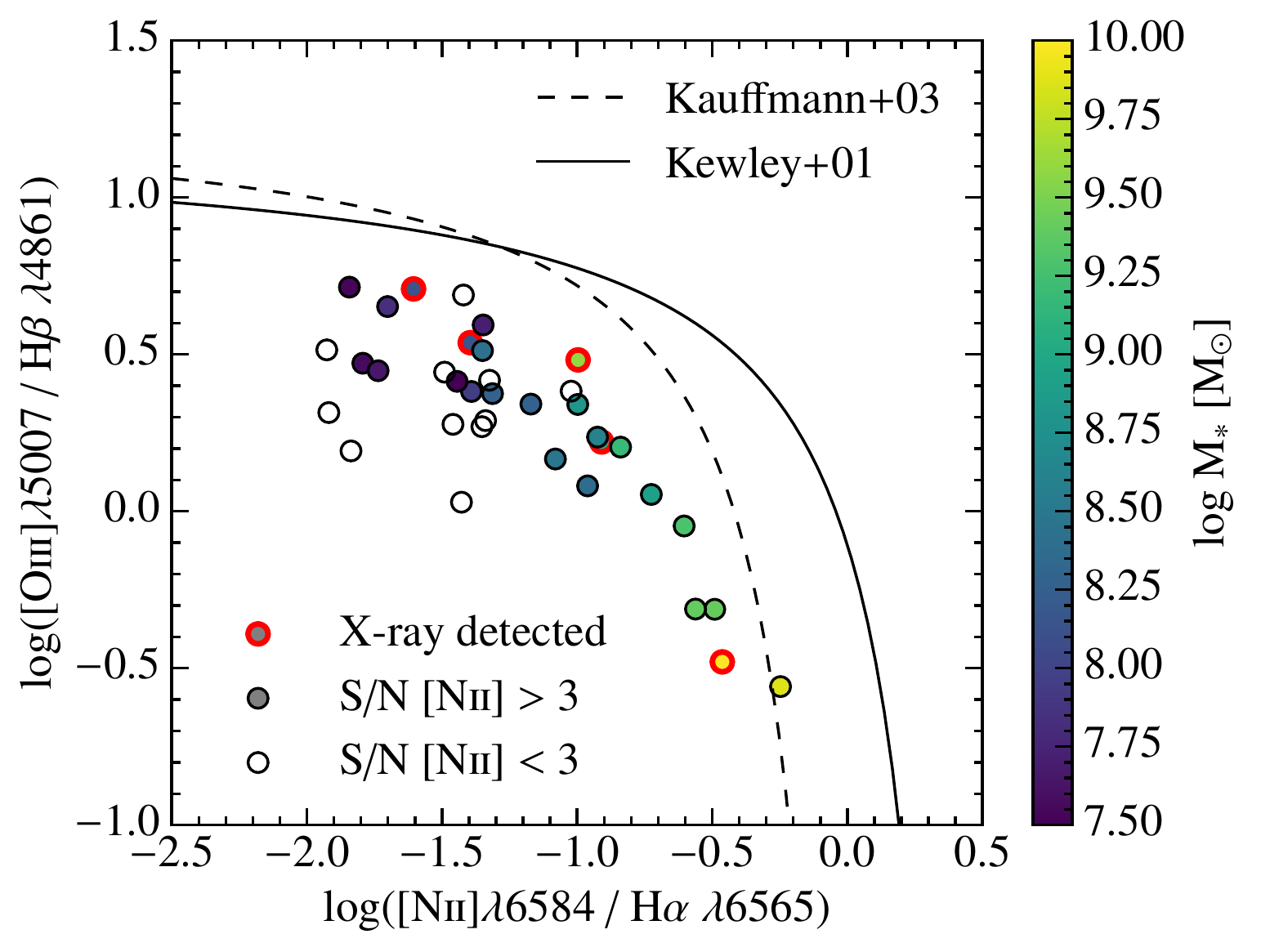}
  \caption{\label{fig:BPT} BPT-diagram \citep{Baldwin1981} of the sources in our
    \Halpha-subsample for which we measure \Nii.  All galaxies fall in the
    star-forming region of the diagram.  The filled and open circles have
    S/N(\Nii) $>3$ and $<3$, respectively, and the \sampleXrayHa\ sources
    encircled in red are detected in X-rays \citep{Luo2016}.  The solid and
    dashed curve show the AGN boundary and maximum starburst line from
    \cite{Kauffmann2003} and \cite{Kewley2001}, respectively.}
\end{figure}

\begin{figure*}
  \centering
  \includegraphics[width=\columnwidth]{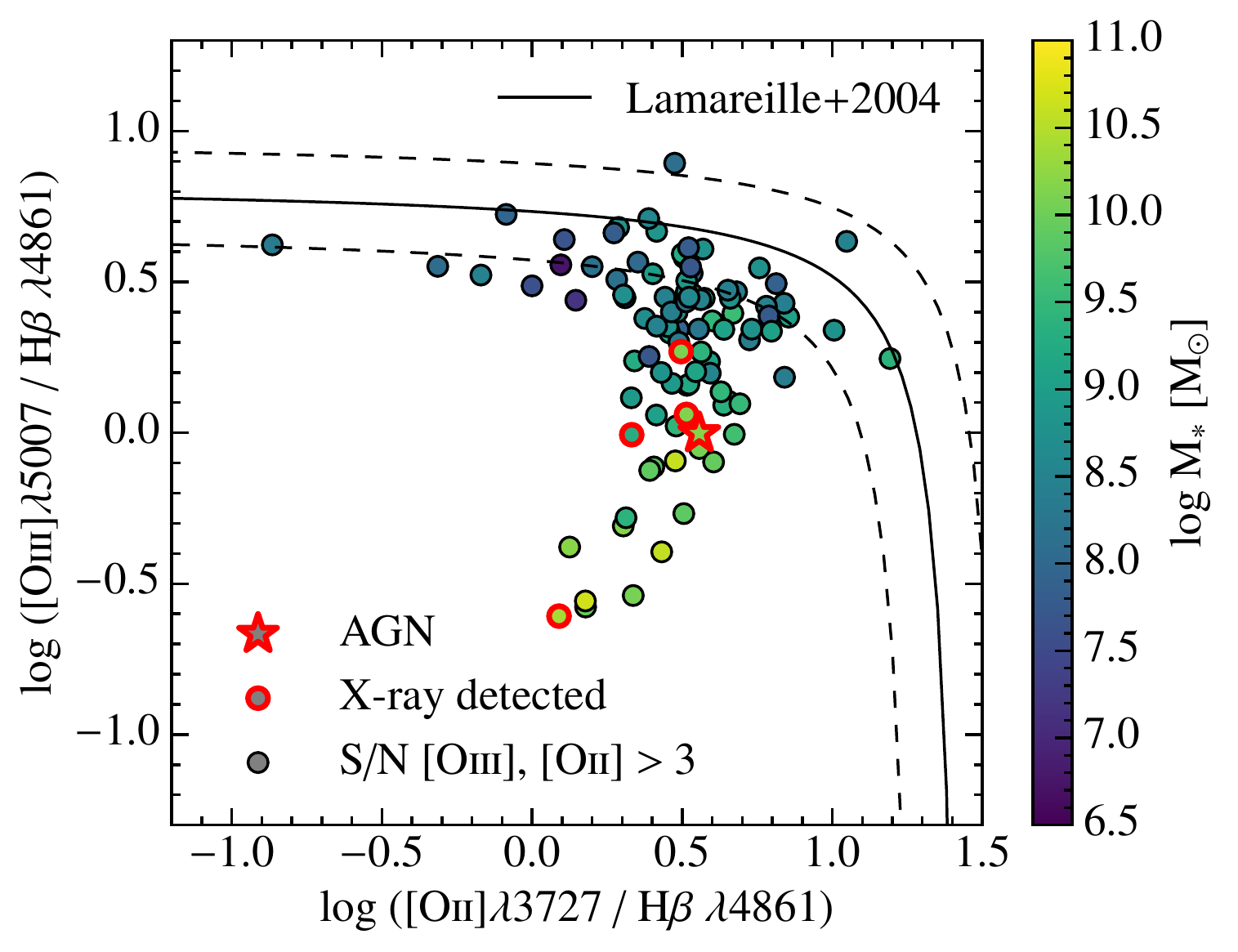}
  \includegraphics[width=\columnwidth]{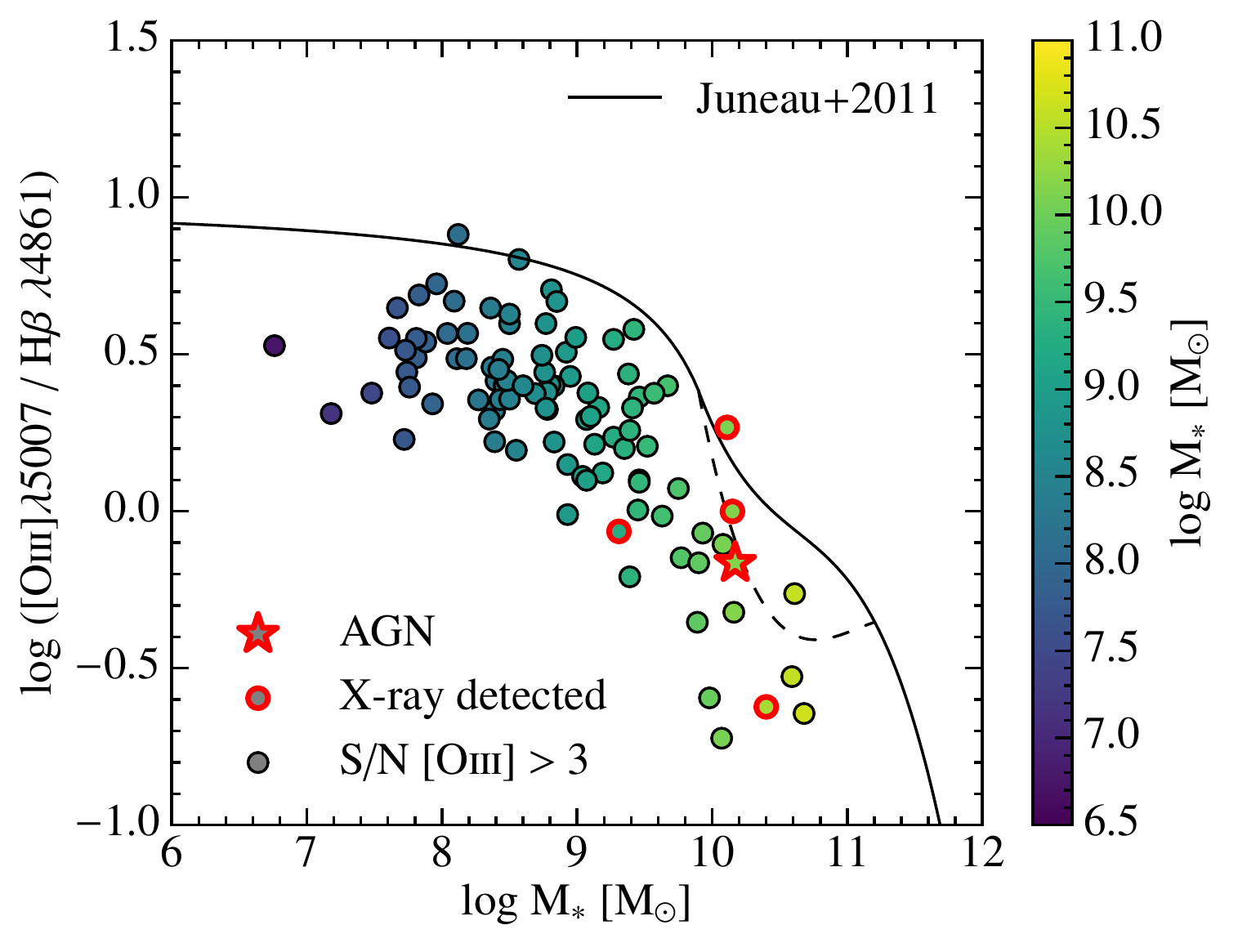}
  \caption{\label{fig:Hb-diag} AGN diagnostics for the sources in our
    \Hbeta-subsample, including all sources which have S/N$>3$ in the relevant
    emission lines.  Overall, our sample is consistent with star-forming
    galaxies. We remove one X-ray detected AGN from the sample.  Left: The
    \Oii/\Hb\ vs. \Oiii/\Hb\ diagnostic from \cite{Lamareille2004} (solid
    line, with the uncertainty indicated by the dashed lines).  Right: The
    mass-excitation diagram from \cite{Juneau2011}.  Galaxies in the region
    between the dashed and solid lines are on average identified as intermediate
    between AGN and SF.}
\end{figure*}

We plot all sources from our \Halpha-subsample for which we have a measurement
of \Nii\ in the BPT-diagram \citep{Baldwin1981} in \Fig{fig:BPT}.  We include
sources for which we have a low S/N (<3) measurement of \Nii\ as open circles.
While we can only put a subsample of our sources on this diagram, all are in the
star-forming region, including the \sampleXrayHa\ galaxies which have an X-ray
detection.  None of the X-ray sources classified as `Galaxy' show spectral
signatures of AGN activity.  In \Fig{fig:Hb-diag} we show a similar consistency
check for the \Hbeta-subsample.  Because we lack access to the BPT diagram at
these redshift, we instead use the diagnostics from both \cite{Lamareille2004}
and \cite{Juneau2011}.  Reassuringly, our sample is overall consistent with
star-forming galaxies and none of the galaxies show line-ratios clearly powered
by AGN activity (including, perhaps surprisingly, the single X-ray classified
AGN).  There is only one source which is above the discriminating line in both
plots (ID\#1114), however, it is consistent within errors with being dominated
by star formation and not detected in X-rays.  Furthermore, its high \OIII\ flux
can very well be driven by star formation and indeed it is part of the sample of
high-\OIII/\OII\ galaxies identified by \cite{Paalvast2018}.  Hence, except for X-ray
detected AGN ID\#875, we do not remove any additional sources from the sample.
Finally, we note that none of the methods to identify AGN are individually
foolproof.  Therefore, we check the impact of potential misclassification of AGN
and confirm that excluding (1) the sources that are above the pure star-forming
line in either of the diagnostic diagrams or (2) all galaxies that are detected
in X-rays (even when consistent with star formation) does not significantly
affect the results.

The final sample then consists of \samplesize\ star-forming galaxies,
\samplehudf\ from the HUDF, all with the highest redshift confidence \citep{Inami2017},
and \samplehdfs\ from the HDFS, between $\samplezmin < z < \samplezmax$ with a mean redshift
of \samplezmean\ (see \Fig{fig:redshift-sfct}).

\subsection{Stellar masses}
\label{sec:stellar-masses}
The stellar masses of the galaxies were estimated using the Stellar Population
Synthesis (SPS) code FAST \citep{Kriek2009}.  The SPS-templates were
$\chi^{2}$-fitted to the broad-band photometry of the different fields for a
range of parameters.  For the HUDF, we rely on the deep \emph{HST} photometry
from the UVUDF catalogue \citep{Rafelski2015} (containing WFC3/UVIS F225W, F275W
and F336W; ACS/WFC F435W, F606W, F775W, and F850LP and WFC/IR F105W, F125W, F140W
and F160W) while for the HDFS we take the WFPC2 photometry from
\cite{Casertano2000} (F330W, F450W, F606W, and F814W).  The SPS-templates were
constructed from the \cite{Conroy2010} (FSPS) models using a discrete range of
metallicities ($Z/Z_{\odot}=[0.04, 0.20, 0.50, 1.0, 1.58]$).  We assumed a
\cite{Chabrier2003} initial mass function with an exponentially declining star
formation history (SFR $\propto \exp(-t/\tau)$ with
$8.5 < \log(\tau/\mathrm{yr}) < 10$ in steps of 0.2 dex).  The redshifts were
fixed to the accurate spectroscopic values determined from the MUSE spectra.
Ages were allowed to vary between $8 < \log \mathrm{Age/yr} < 10.2$ in steps of
0.2 dex.  We parameterised the dust attenuation curve according to the
\cite{Calzetti2000} dust law with the dust extinction in the visual taken to be
within $0 < A_{V} < 3$ ($\Delta A_{V} = 0.1$ magnitudes).  For all the
parameters error estimates were obtained through Monte Carlo methods, by
re-running the fitting 500 times while varying the input photometry within their
photometric errors (see \citealt{Kriek2009} for details).

\begin{figure}
  \centering
  \includegraphics[width=\columnwidth]{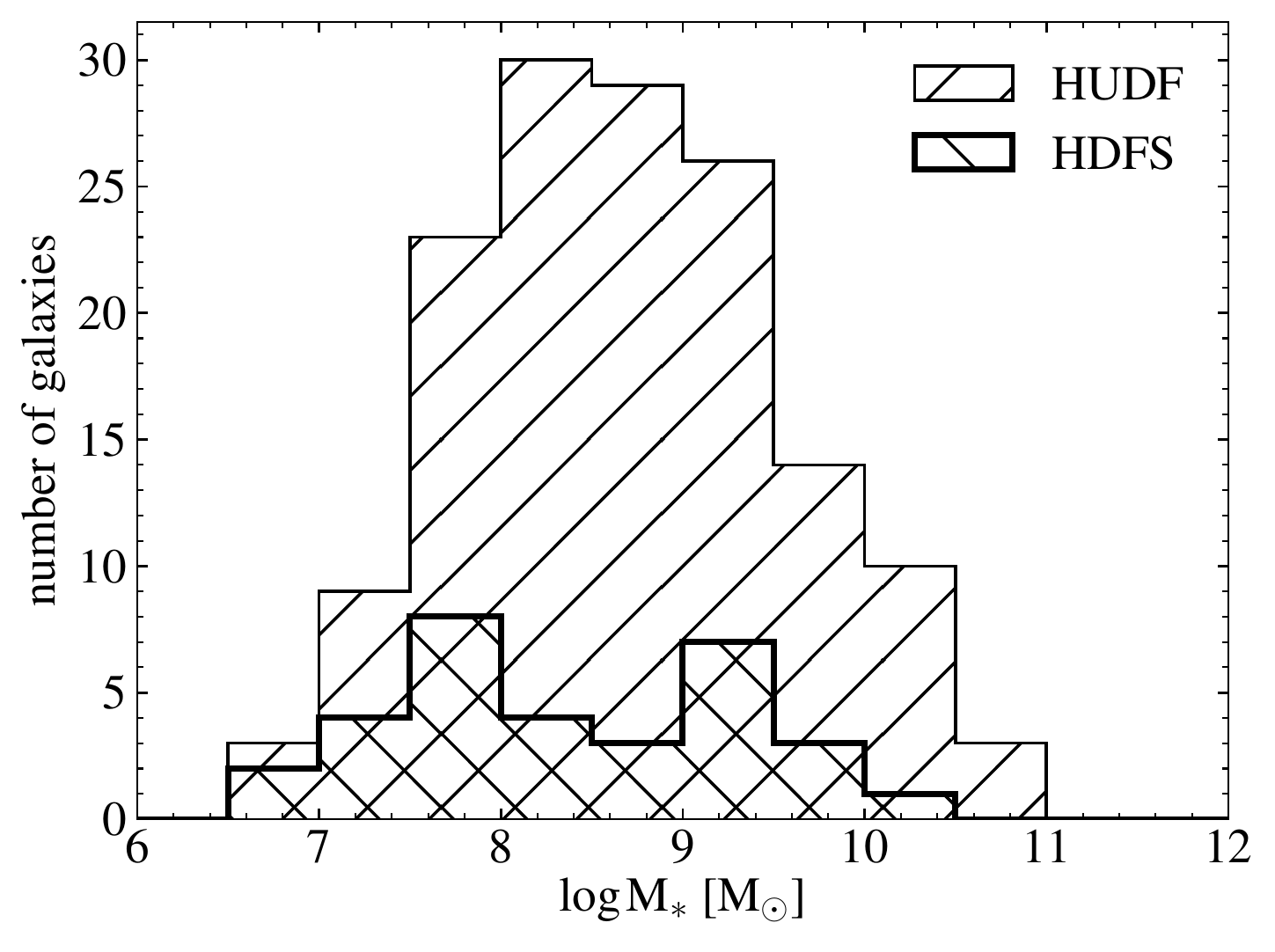}
  \caption{\label{fig:mass-histogram} Histograms of the stellar mass
    distributions of the MUSE detected galaxies in the HUDF and the HDFS.  The
    deep 30h observations allow us to detect and subsequently infer a stellar
    mass and SFR for galaxies down to $\sim \SI{e7}{\solarmass}$.}
\end{figure}

Stellar masses were determined for all \samplesize\ objects in the final sample.
The distribution of masses is shown in \Fig{fig:mass-histogram}.  With these
deep MUSE observations we are mainly probing low-mass (\SI{<e9.5}{\solarmass})
galaxies and we can still detect star formation from emission lines in galaxies
with mass $\sim$\SI{e7}{\solarmass}.  The mass estimates with their upper and
lower confidence intervals are shown for the individual objects in
\Fig{fig:m*-sfr}.  The mean and standard deviation of the average errors on the
mass estimates are $0.19 \pm 0.06$ dex for the HUDF and $0.22\pm0.12$ dex for
the HDFS.

\subsection{Star formation rates}
\label{sec:star-formation-rates}
The star formation rates are inferred from the flux in the \Halpha\ or \Hbeta\
recombination lines emitted by \HII\ regions, which primarily trace recent
(\SI{\sim 10}{\mega\year}) massive star formation.  Before we can infer a SFR we
need to correct the measured flux in the emission lines for the attenuation by
dust along the line of sight.  We do this assuming a dust law according to
\cite{Charlot2000} (i.e. $\tau \propto \lambda^{-1.3}$, appropriate for birth
clouds) and using the intrinsic ratio of the Balmer recombination lines
($j_{\Ha}/j_{\Hb} = 2.86$ and $j_{\Hb}/j_{\Hg} = 2.14$; \cite{Hummer1987}, for
an electron temperature and density of $T = \SI{10000}{\kelvin}$ and
$n_{e} = \SI{e3}{cm^{-3}}$).  Hence, to derive an SFR(\Halpha) we also require a
measurement of \Hbeta\ and likewise for SFR(\Hbeta) we also require \Hgamma.
After the dust correction we can convert the intrinsic flux to a luminosity
using the measured redshift, given the assumed $\Lambda$CDM cosmology.

To determine the SFR we follow the treatment by \cite{Moustakas2006}, which is
essentially based on the relations from \cite{Kennicutt1998}.  Out of the SFR
indicators that MUSE has access to, the \Halpha\ line presents the least
systematic uncertainties, but it is only available at low redshifts
($z \la 0.42$ for MUSE at \SI{9300}{\angstrom}; \sampleHalpha\ galaxies).  We
convert the \cite{Kennicutt1998} relation from a Salpeter to a Chabrier IMF
($0.1 < M [\si{\solarmass}] < 100$) by multiplying by a factor 0.62 (which is
derived by computing the difference in total mass in both IMFs, while assuming
the same number of massive (\SI{>10}{\solarmass}) stars):
\begin{align}
  \label{eq:sfr-ha}
  \mathrm{SFR}(\Halpha) = 4.9 \times 10^{-42}\frac{L(\Halpha)}{\si{erg s^{-1}}}\si{\solarmass \per \year},
\end{align}
where $L(\Halpha)$ is the dust-corrected luminosity.  We note that this
calibration assumes case B recombination and solar metallicity.

Because \Halpha\ moves out of the optical regime at $z>0.42$, the \Hbeta\
luminosity is the primary tracer of SFR for the majority of our sample
(\sampleHbeta\ galaxies).  Given the intrinsic flux ratio between \Halpha\ and
\Hbeta, we can convert equation \Eq{eq:sfr-ha} into a SFR for $L(\Hbeta)$:
\begin{align}
  \label{eq:sfr-hb}
  \mathrm{SFR}(\Hbeta) = 1.4 \times 10^{-41} \frac{L(\Hbeta)}{\si{erg s^{-1}}}\si{\solarmass \per \year},
\end{align}
where $L(\Hbeta)$ is corrected for dust.  We note that the \Hbeta\ derived SFR
inherits all the uncertainties from SFR$(\Halpha)$, including variations in dust
reddening \citep{Moustakas2006}.

We also investigate the SFR using the \Oii\ nebular emission line.  Here we use
the calibration for the \Halpha\ SFR (\Eq{eq:sfr-ha}), where we assume an
intrinsic flux ratio between \Oii\ and \Halpha\ of unity \citep{Moustakas2006}.
Since \Oii\ is closest to \Hbeta, we use the \Hbeta/\Hgamma\ ratio to determine
the dust correction, scaled to the appropriate wavelength.  The consequence of
this is that the addition of the \Oii\ line as a tracer of SFR will not add any
new objects to the sample.  Instead, it can be used as a useful comparison,
which will be discussed in \Sec{sec:cons-sfr-indic}.

To estimate the uncertainty in the SFR estimates (and dust corrections), we use
Monte Carlo methods to derive a confidence interval on the SFR of every
individual galaxy.  We create a posterior distribution on the SFR by doing 1000
draws from a Gaussian distribution centred on the measured flux, with the
variance set by the measurement error squared.  The median posterior SFR can
then be determined, as well as the $\pm 1 \sigma$ confidence intervals, by
taking the $50^{\text{th}}$, $16^{\text{th}}$ and $84^{\text{th}}$ percentile
from the derived posterior distribution.

\begin{figure*}
  \centering
  \includegraphics[width=\textwidth]{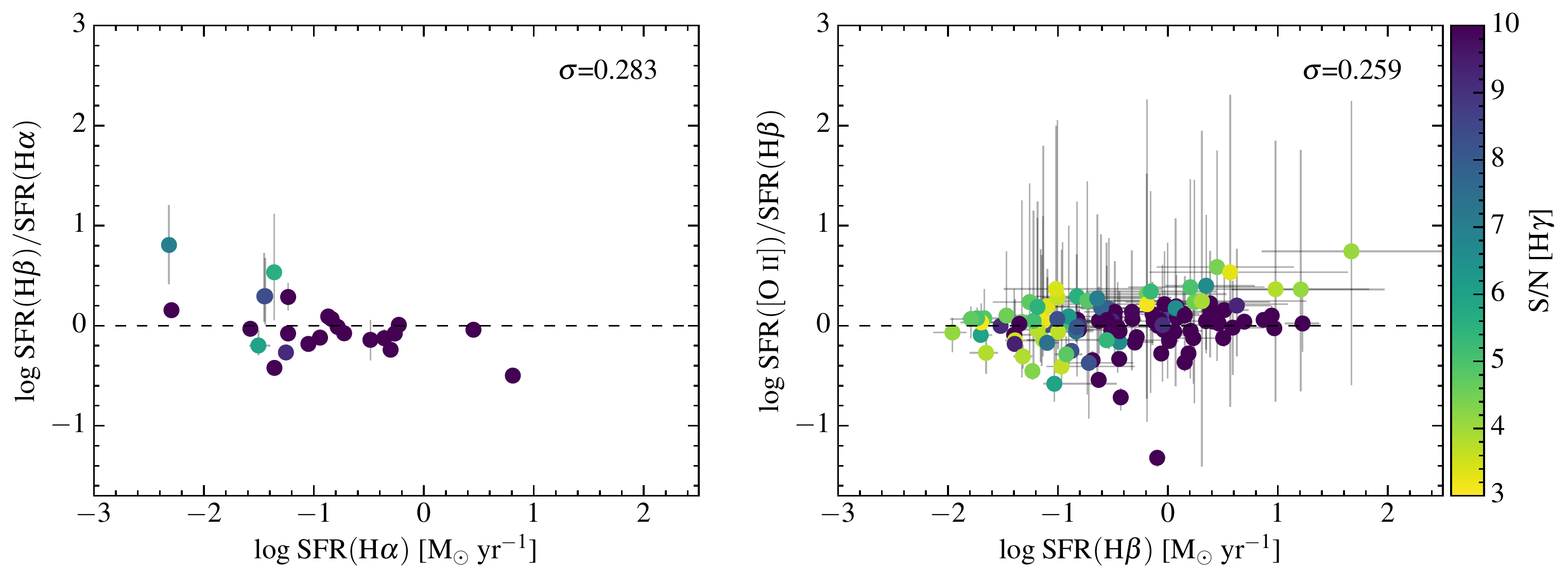}
  \includegraphics[width=\textwidth]{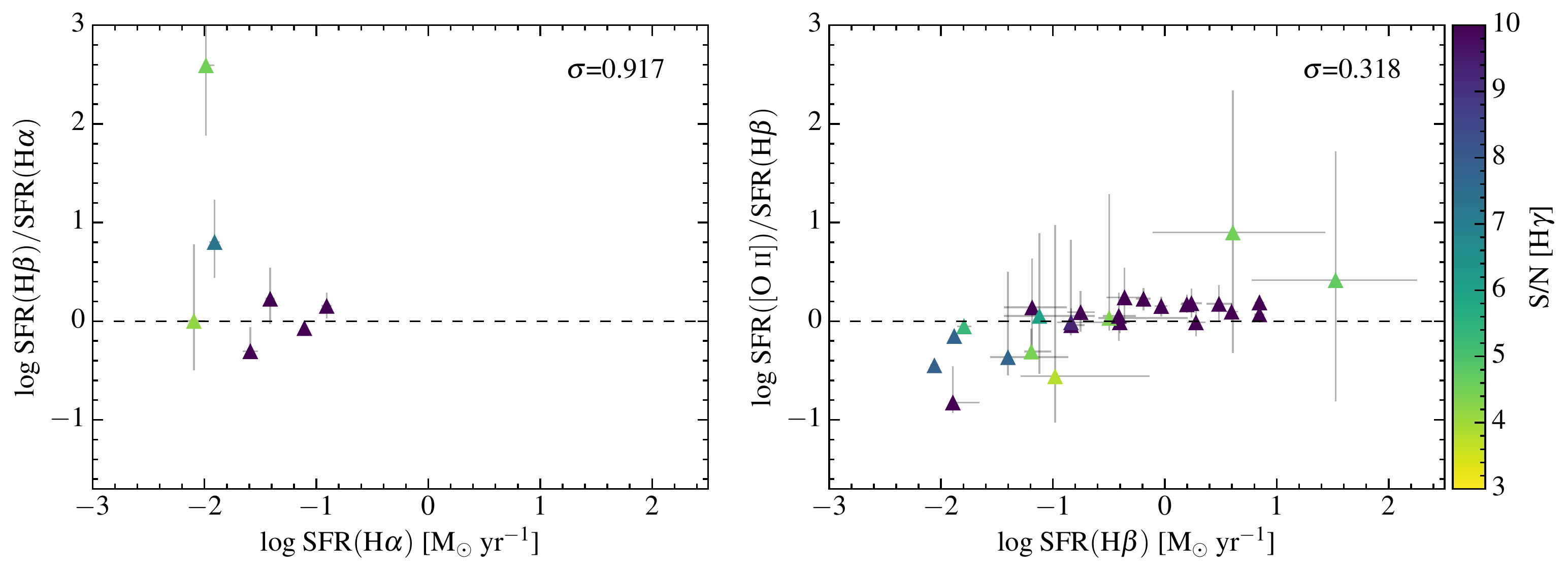}
  \caption{\label{fig:sfr-consistency} A comparison of the derived star
    formation rate (SFR) from the \Halpha, \Hbeta\ and \Oii\ luminosities for
    the HUDF (top panels, circles) and the HDFS (bottom panels, triangles). The
    left panels show the logarithm of the SFR from \Halpha\ vs. the
    difference between the log \Hbeta\ and log \Halpha\ SFRs.  The right panels
    show the same for \Hbeta\ vs. \Oii.  In the top right corners $\sigma$
    indicates the standard deviation (in dex) around the one-to-one relation.
    Colour indicates the signal-to-noise ratio (S/N) in the faintest line;
    \Hgamma.  Only galaxies that allowed for more than one SFR indicator are
    included in the plot.  Overall the SFRs from \Hbeta\ and \Oii\ agree
    reasonably well, considering we have not taken into account the metallicity
    dependence in SFR$(\Oii)$.  The scatter in \Halpha\ vs. \Hbeta\ SFR is
    largely driven by \Hgamma\
    S/N.}
\end{figure*}

\begin{figure}
  \centering
  \includegraphics[width=\columnwidth]{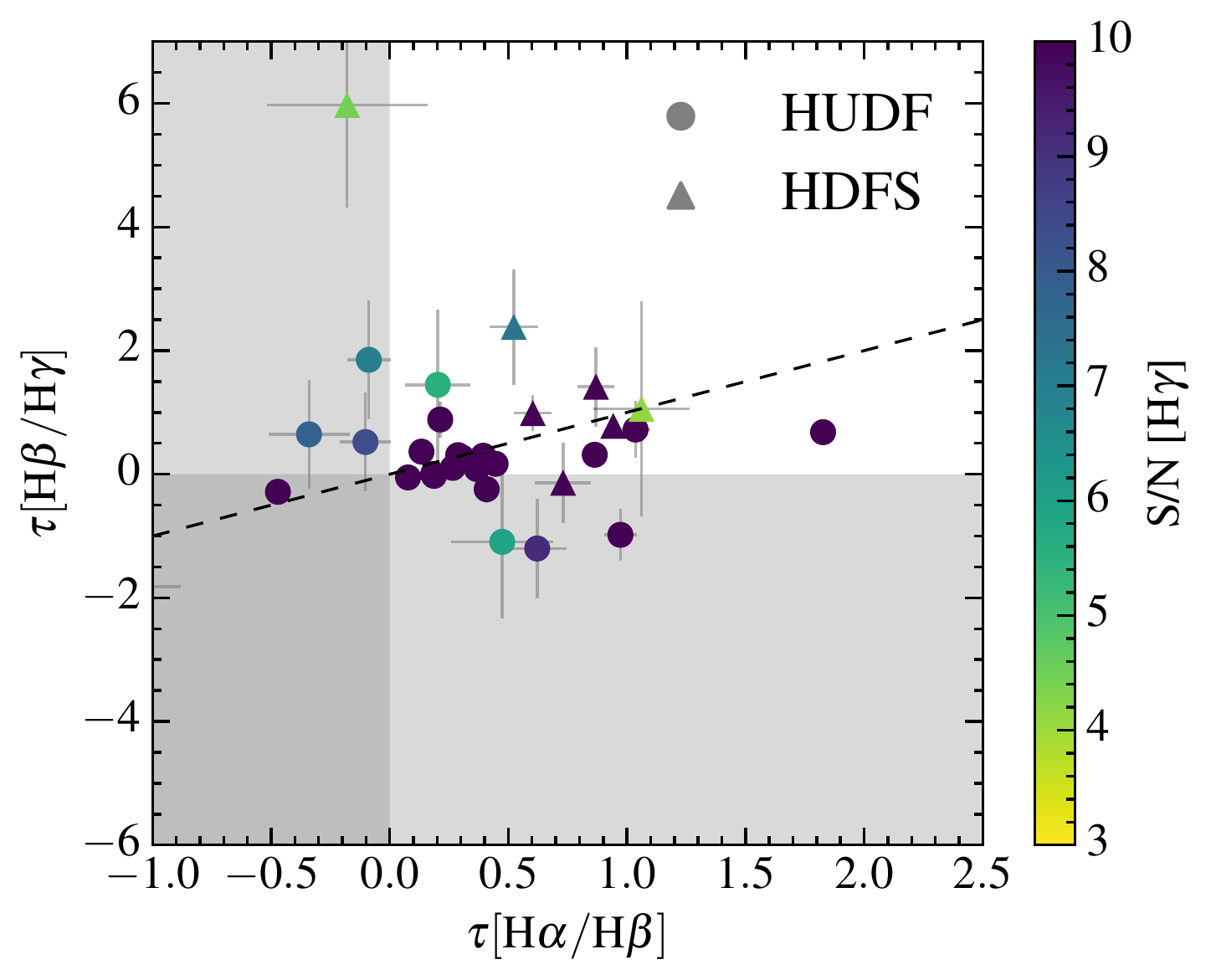}
  \caption{\label{fig:tau-comparison} Optical depths at the wavelength of
    \Hbeta\ as derived from both the \Hbeta/\Hgamma\ and the
    \Halpha/\Hbeta-ratio, coloured by \Hgamma\ signal-to-noise (S/N).  The
    dashed line is the one-to-one relation.  Overall the optical depths agree
    reasonably well, unless the \Hg\ S/N is low.  Most galaxies actually show
    little dust ($\tau$ close to zero).  The shaded area shows the regions of
    (unphysical) negative optical depth for each axis.  We set the optical
      depth to zero for galaxies with negative $\tau$ as this is often
      consistent with the error bars and the offset is due to noise in the
      spectra.  We note that some of the high-S/N outliers actually have
    discrepant Balmer ratios.  If the inferred optical depth is very different,
    this will affect the comparison of the dust-corrected SFR from \Hbeta\ and
    \Halpha\ (see \Fig{fig:sfr-consistency}).}
\end{figure}

\section{Consistency of SFR indicators}
\label{sec:cons-sfr-indic}
Before turning to the results, we first consider the consistency of the derived
SFRs, by comparing the SFR estimates from different tracers for the same
galaxies.  In the remainder of the paper we only use the dust-corrected Balmer
lines as tracers of star formation.

For a significant fraction of our galaxies ($\approx 40 \%$) we find that the
Balmer line ratios are below their case B values (as stated in
\Sec{sec:star-formation-rates}), indicative of a negative dust correction.
While this might seem surprising, this is not uncommon and similar ratios have
been seen in spectra from, e.g. the SDSS \citep{Groves2012}, MOSDEF
\citep{Reddy2015}, KBSS \citep{Strom2017} and ZFIRE \citep{Nanayakkara2017}.
While `unphysical', these ratios are not entirely unexpected and can have
several causes.

First, these deviations can be caused by noisy spectra.  Most galaxies in our
sample are not very dusty and hence have a ratio close to case B.  In $> 50 \%$
of the cases with deviant ratios, the case B ratio is indeed within the
$1\sigma$ error bars.  We conservatively apply no dust correction for all these
galaxies.  The mean dust correction for all galaxies in our sample is
$\tau(\Hb/\Hg) \approx 0.6$ (setting galaxies with a negative dust correction to
zero) or $\tau(\Hb/\Hg) \approx 1$ (including only galaxies with a positive dust
correction).

Secondly, there could be a problem with the measurement.  Three objects that
were significantly offset from the rest of the sample showed particular problems
in their emission lines.  In one object (ID\#971) \Hgamma\ was severely affected
by an emission line from a nearby source (\Oiiia\ from ID\#874 at $z=0.458$,
another galaxy in our sample, coincidentally almost exactly at the observed
wavelength of \Hg).  For five other objects there was a clear problem with the
fit to the \Hbeta\ (ID\#894, \#896, \#1027) or \Halpha\ (ID\#2, \#1426) emission
lines.  We subsequently removed the first four sources from the analysis; for
the latter two we disregarded the \Halpha\ SFR and use the \Hbeta\ SFR.

A third, intriguing option is that theses objects are real.  Indeed, there
remains a small number of galaxies which have high-S/N spectra, but still show
Balmer ratio's below their case B values.~\footnote{ It is important to point
  out that this is not caused by stellar absorption in the continuum as this is
  taken into account when modelling the emission lines with \textsf{platefit}.}.
Similar objects have also been observed in the other surveys already referenced,
such as SDSS (Jarle Brinchmann, private communication, see also
\citealt{Groves2012}).  While these are very interesting objects on their own, a
detailed analysis of these sources is beyond the scope of this paper.  To be
conservative and consistent, we apply no dust correction for these sources.

For some objects in the sample we measure multiple emission lines, which allows
us to infer a SFR from different tracers.  In any case a pair of Balmer lines
(either \Ha/\Hb\ or \Hb/\Hg) is available (\Sec{sec:sample-selection}), to allow
for a dust correction.  The majority of our sample lies at $z > 0.42$ for which
\Halpha\ is not available, but (dust-corrected) \Oii\ is available as an SFR
indicator.  In \Fig{fig:sfr-consistency} we show a comparison for all galaxies
that allowed both \Halpha\ and \Hbeta\ (only some galaxies at $z < 0.42$) and
\Hbeta\ and \Oii\ derived SFRs (all redshifts).  We note that \Hbeta\ and \Oii\
derived SFRs are corrected for dust using the same \Hb/\Hg-ratio.

In the right panels of \Fig{fig:sfr-consistency} we see that the \Hbeta\ and
\Oii\ derived SFRs agree remarkably well (standard deviation $\sigma \leq 0.28$
dex), considering that we have not taken into account the metallicity dependence
of the \Oii\ luminosity in the SFR conversion factor \citep[e.g.][]{Kewley2004}.
A few points scatter quite a bit, most of which have large error bars.  At lower
SFRs we do see that \Oii\ predicts a lower SFR than \Hbeta, which is probably
because at low SFR we are also probing low-mass and low-metallicity galaxies.
Stars with a lower metallicity have a higher UV flux, which causes the
ionisation equilibrium for oxygen to shift from \OII\ to \OIII\, which
diminishes the observed \Oii\ flux.  Because of the opposite effect \Oii\
occasionally predicts a higher SFR than \Hbeta\ at the high-SFR end.

For a limited number of objects all three Balmer lines are in the spectral range
of MUSE ($0.09 < z < 0.42$).  We compare the \Halpha\ and \Hbeta\ derived SFRs
in the left panel of \Fig{fig:sfr-consistency}, where we find reasonable
agreement (in the HUDF, where we have most sources, they have a factor of
$\sim 2$ scatter).  Most of the scatter is found at low SFR, where (on average)
the S/N is also the lowest.  In the HDFS one object (at low S/N) is a strong
outlier, but removing this source yields a similar scatter to the HUDF.
Intuitively the SFRs from \Ha\ and \Hb\ should agree very well, which warrants
some deeper investigation into the outliers at low SFR.

The main uncertainty in the SFR estimate is the amount of dust attenuation.  In
\Fig{fig:tau-comparison} we compare the inferred optical depth from the
\Hb/\Hg-ratio ($\tau[\Hb/\Hg]$) to the optical depth determined from the
\Ha/\Hb\ ratio ($\tau[\Ha/\Hb]$).  We note though that \Fig{fig:tau-comparison}
shows the measured optical depth, while we set negative $\tau$ to zero before
computing the SFR.  Indeed, while many sources agree well, we see that the
amount of dust correction estimated from the Balmer lines is not consistent for
several objects, leading to a different SFR estimate from \Halpha\ and \Hbeta.

This tension is in part caused by the nature of the experiment, which requires
that all three Balmer lines are in the spectral range of MUSE simultaneously.
Necessarily then, \Halpha\ will be at the long wavelength end of the
spectrograph where skylines are more prevalent, occasionally adding uncertainty
to its measurement.  For the low-SFR sources, however, \Hgamma\ might not be
very bright, adding uncertainty to the dust correction of SFR$({\Hbeta})$ for
these sources (as seen at lower SFR in the left panels of
\Fig{fig:sfr-consistency}).  Indeed, most of the outliers have a low S/N in
\Hgamma\ (as stated earlier, for the objects with a negative dust correction
from \Hb/\Hg, we leave the often lower S/N measurement of \Hgamma\ out of the
analysis by setting $\tau(\Hb/\Hg)$ to zero).  On the other hand, the converse
is not quite true: for a large number of sources with a low S/N in \Hgamma\ we
do have a consistent SFR estimate.  For all objects we use the highest S/N lines
available to infer a dust-corrected SFR, i.e. for objects which have a
measurement of all three Balmer line we use the \Halpha, \Hbeta\ pair to infer a
dust-corrected SFR, which generally has the highest S/N.

In summary, we have dust-corrected SFR measurement from the \Halpha\ and \Hbeta\
spectral lines for all galaxies at $z < 0.42$ and the \Hbeta, \Hgamma-pair at
higher redshifts.  Comparing \Halpha\ and \Hbeta\ SFRs, we conclude that the
dust correction is the largest uncertainty on the derived SFR.  We always use
the highest S/N line-pair available to compute a dust-corrected SFR.  Comparing
the \Hbeta\ SFRs with \Oii\ at all redshifts, we see a very consistent picture
(they have $\leq 0.3$ dex scatter in both fields).  Naturally, some variations
between \Hbeta\ and \Oii\ SFRs are expected given the metallicity dependent
nature of \Oii.

\section{Bayesian model}
\label{sec:modelling}
\subsection{Definition}
\label{sec:definition}
The star formation sequence is commonly described by a power-law relation
between stellar mass ($M_{*}$) and star formation rate (SFR), which evolves with
redshift ($z$):
\begin{align}
  \label{eq:powerlaw}
  \mathrm{SFR} \propto M_{*}^{a} (1+z)^{c},
\end{align}
where $a$ and $c$ are the power law exponents.  It has been suggested that the
slope ($a$) becomes shallower in the high-mass regime
($M_{*} > \SI{e10}{\solarmass}$).  In this work we will focus on the low-mass
regime, for which we assume the slope is constant with mass.  We will revisit
this assumption in \Sec{sec:low-mass-sample}.  Given the lack of homogeneous
studies with redshift it is still unclear whether the low-mass slope of the
relation evolves with redshift.  Here, we assume that the low-mass slope is
independent of redshift over the range that we probe in this study.  Likewise,
given the large uncertainties in (the evolution of) the intrinsic scatter, we
limit the number of free parameters in the model and assume that the intrinsic
scatter does not depend on any of the other model parameters.

Following this description, we model the star formation sequence by a plane in
($\log M_{*}$, $\log(1+z)$, $\log{\mathrm{SFR}}$)-space:
\begin{align}
  \label{eq:linear}
  \log{\mathrm{SFR}[\si{\solarmass\per\year}]} =  a \log{\left(\frac{M_{*}}{M_{0}}\right)} +b + c\log{\left(\frac{1+z}{1+z_{0}}\right)},
\end{align}
where $b$ is now a normalisation constant.  We take
$M_{0} = \SI{e8.5}{\solarmass}$ and $z_{0} = 0.55$ (close to the medians of the
data) without the loss of generality.  Galaxies scatter around this relation
with an amount of intrinsic scatter in the vertical (i.e. $\log{\mathrm{SFR}}$)
direction, which we denote by $\sigma_{\mathrm{intr}}$.  In the lack of an
obvious alternative, we take the intrinsic scatter to be Gaussian in our model.

In a statistical sense we can then say that our observations ($\log M_{*}$,
$\log(1+z)$, $\log \text{SFR}$) are drawn from a Gaussian distribution around
the plane defined by \Eq{eq:linear}.  To recover this distribution, we need to
take a careful approach, taking into account the heteroscedastic errors of the
measurements.

We adopt a Bayesian approach to determine the posterior distribution of the
model parameters ($a, c, b, \sigma_{\mathrm{intr}}$) (see \cite{Andreon2010} for
a lucid description of the Bayesian methodology in an astronomical context).
Different approaches to construct the likelihood have been presented in the
literature (see e.g. \citealt{Kelly2007} or \citealt{Hogg2010}).  We choose to
adopt a parameterisation of the likelihood following \cite{Robotham2015}
(hereafter \RO).

First, we state that our knowledge about galaxy $i$ (determined by the
observations) is encompassed by the probability density function of a
multivariate Gaussian distribution, $\mathcal{N}(\mathbf{x}_{i}, C_{i})$, with a
mean value of:
\begin{align}
  \label{eq:mean}
  \mathbf{x}_{i} = (\log M_{*,i}, \log(1+z_{i}), \log \mathrm{SFR}_{i})
\end{align}
and a diagonal covariance matrix:
\begin{align}
 \label{eq:cov}
 C_{i} =
  \begin{pmatrix}
    \sigma_{\log M_{*},i}^{2} & 0 & 0 \\
    0 & \sigma_{\log (1+z),i}^{2} & 0  \\
    0 & 0 & \sigma_{\log \mathrm{SFR}, i}^{2}
  \end{pmatrix}
\end{align}
containing the variance in each parameter.  This is justified as both stellar
mass and star formation rate are measured independently from different data.
The covariance with redshift is negligible as the error on the spectroscopic
redshift is very small.

Secondly, we parameterise the model given by \Eq{eq:linear} (which is a plane in
three dimensions) in terms of its normal vector $\mathbf{n}$, to avoid
optimisation problems \citepalias{Robotham2015}.  The galaxies scatter around
this plane with an amount of intrinsic Gaussian scatter, perpendicular to the
plane, which we denote by $\sigma_{\perp}$.  We note that perpendicular scatter
$\sigma_{\perp}$ is distinct from the (commonly reported) vertical scatter
$\sigma_{\mathrm{intr}}$ which lies in the $\log \mathrm{SFR}$ direction.  After
the analysis, we can simply transform the parameters
($\mathbf{n}, \sigma_{\perp}^{2}$) back into familiar parameters
$(a,c,b,\sigma_{\mathrm{intr}}^{2})$ (using \citetalias{Robotham2015}, Eq. 9).

Given the above definitions, we can express our
log-likelihood~\footnote{Throughout this paper we consistently use `$\log$' for
  the base-10 logarithm and `$\ln$' for the base-$e$ logarithm, with one
  exception: we stick to standard terminology and call $\ln{\mathcal{L}}$ the
  `log-likelihood'.} as the sum over $N$ data points (see also
\citetalias{Robotham2015}):
\begin{align}
  \label{eq:likelihood}
  \ln \mathcal{L} = -\frac{1}{2}\sum_{i=1}^{N} \left[\ln \left(\sigma_{\perp}^{2} + \frac{\mathbf{n}^{\top} C_{i}\mathbf{n}}{\mathbf{n}^{\top}\mathbf{n}}\right)
  + \frac{(\mathbf{n}^{\top}[\mathbf{x}_{i} - \mathbf{n}])^2}{\sigma_{\perp}^{2} \mathbf{n}^{\top}\mathbf{n} + \mathbf{n}^{\top}C_{i}\mathbf{n}} \right],
\end{align}
where all the parameters have been defined earlier.

Lastly, we have to define our priors on each component of $\mathbf{n}$ and on
$\sigma_{\perp}^{2}$.  As we want to impose limited prior knowledge, we express
our priors as uniform distributions, with large bounds compared to the typical
values of the parameters (we confirm that the results are robust, irrespective
of the exact choice of bounds).
\begin{align}
  \label{eq:prior}
  \mathbf{n} &\sim \mathcal{U}^{3}(-1000,1000) \\
  \sigma_{\perp}^2 &\sim \mathcal{U}(0,1000), \notag
\end{align}
where $\mathcal{U}^{n}$ is the $n$-dimensional multivariate uniform distribution
and we take into account the fact that variance is always positive.

\subsection{Execution}
\label{sec:execution}
With the likelihood and priors in hand we determine the posterior using
\emph{Markov chain Monte Carlo} (MCMC) methods.  We use the Python
implementation called \textsf{emcee} \citep{Foreman-Mackey2013}, which utilises
the affine-invariant ensemble sampler for MCMC from \cite{Goodman2010}.
\textsf{emcee} samples the parameter space in parallel by setting off a
predefined number of `walkers', which we take to be 250.

Following \cite{Foreman-Mackey2013}, we first initialise the walkers randomly in
a large volume of parameter space.  We then restart the walkers in a small
Gaussian ball around the median of the posterior distribution (i.e. around the
`best solution').  We (generously) burn in for a quarter of the total amount of
iterations for each walker which we take to be 20000 for the main run
(\Sec{sec:global-sample}; roughly four hundred times the autocorrelation time).
We note that for all subsequent runs described below we follow the same
procedure, with similar results.

We take several steps to check whether the \textsf{emcee} algorithm has properly
converged.  As an indication, one can look at both the mean acceptance fraction
of the samples as well as the autocorrelation time \citep{Foreman-Mackey2013}.
For the main run the acceptance fraction that resulted from the modelling (0.45)
was well within range advocated by \cite{Foreman-Mackey2013} (0.2 - 0.5).  The
autocorrelation time was also relatively short and we let the walkers sample the
posterior well over the autocorrelation time.  Furthermore, we confirmed that
the walkers properly explored the parameter space.

Combining the results from all walkers then gives the posterior distribution
over which we can marginalise to find the posterior probability distributions
for the model parameters.  We will discuss the results of the modelling in
\Sec{sec:results}.

\subsection{Model and data limitations}
\label{sec:model-data-lims}
The unique aspect of the likelihood in \Eq{eq:likelihood} is that it captures
both the heteroscedastic errors on the observables as well as the intrinsic
scatter around the plane.  Furthermore, it can simultaneously describe both the
slope of the sequence as well as the evolution with redshift.

It is important to determine how well we can recover the `true' parameters with
the observed data at hand.  Our MUSE observations are constrained by the fact
that we can only detect galaxies in a certain redshift range and cannot
detect galaxies below the flux limit of the instrument (see
\Fig{fig:redshift-sfct}).  As the flux limit varies with redshift, this could
introduce a bias in our inferred parameters.  The reason behind this is that the
lack of low-SFR galaxies at higher redshift will bias the posterior towards
shallower slopes, with a steeper redshift evolution (see \Fig{fig:sfc} for an
illustration).  In order to correct for such a bias, we analyse a series of
simulated observations.  We briefly outline the procedure here, which is
described in detail in \autoref{sec:simulations}.

In order to characterise the bias in the inferred parameters, we simulate
galaxies from a mock star formation sequence for a range of values in each
parameter, which we call $\mathbf{x}_{\mathrm{true, k}}$ (see
\autoref{tab:mock-grid-input}).  After applying the redshift-dependent flux
limit to the mock data, we model the remaining galaxies as described in
\Sec{sec:modelling} and recover the parameters, $\mathbf{x}_{\mathrm{out, k}}$.
We then fit the transformation between the true and recovered parameters with an
affine transformation
($\mathbf{x}_{\mathrm{out, k}} = A \mathbf{x}_{\mathrm{true, k}} + \mathbf{b}$)
as outlined in \Sec{sec:line-transf}.  The inverse of the best-fit
transformation (\Eq{eq:inv-trans}) can then be used to correct the posterior
density distribution as measured from the MUSE data.  In the following, we
provide both the uncorrected (directly fitted) and the corrected values for
reference.

\section{Star formation sequence}
\label{sec:results}
\begin{figure*}
  \centering
  \includegraphics[width=\textwidth]{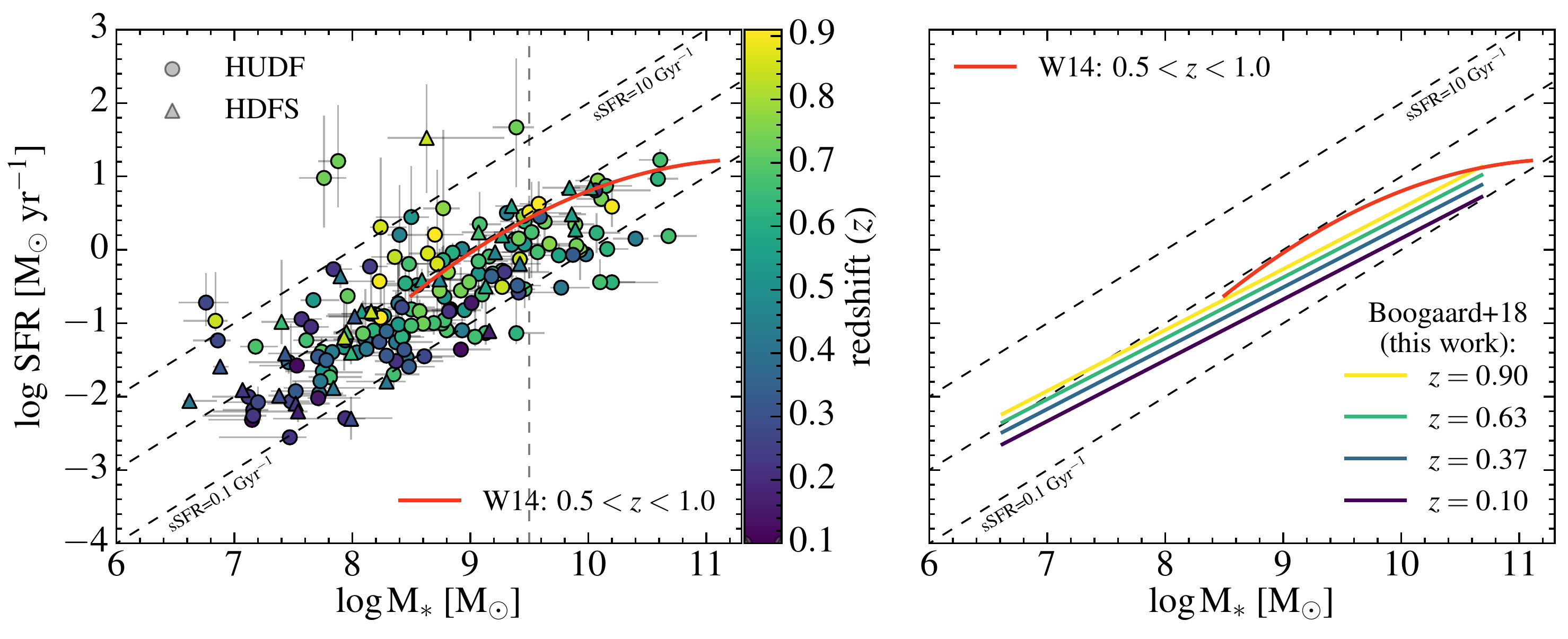}
  \caption{\label{fig:m*-sfr} Left panel: The sample of \samplesize\
    star-forming galaxies observed with MUSE, plotted on the $M_{*}$-SFR plane.
    The symbols indicate the field and colour indicates the redshift.  The
    dashed lines show a constant sSFR, which is equivalent to a linear
    relationship: $\mathrm{SFR} \propto M_{*}$.  The red curve shows the model
    of the star formation sequence from \protect\cite{Whitaker2014} for
    $0.5 < z < 1.0$.  The vertical grey dashed line indicates the selection for
    the low-mass fit (\Sec{sec:low-mass-sample}).  Right panel: Same as the left
    panel but with the data points removed, showing (the evolution of) the star
    formation sequence as seen by MUSE, according to \Eq{eq:best-fit-corr}.}
\end{figure*}

\begin{figure*}
  \centering
  \includegraphics[width=0.7\textwidth]{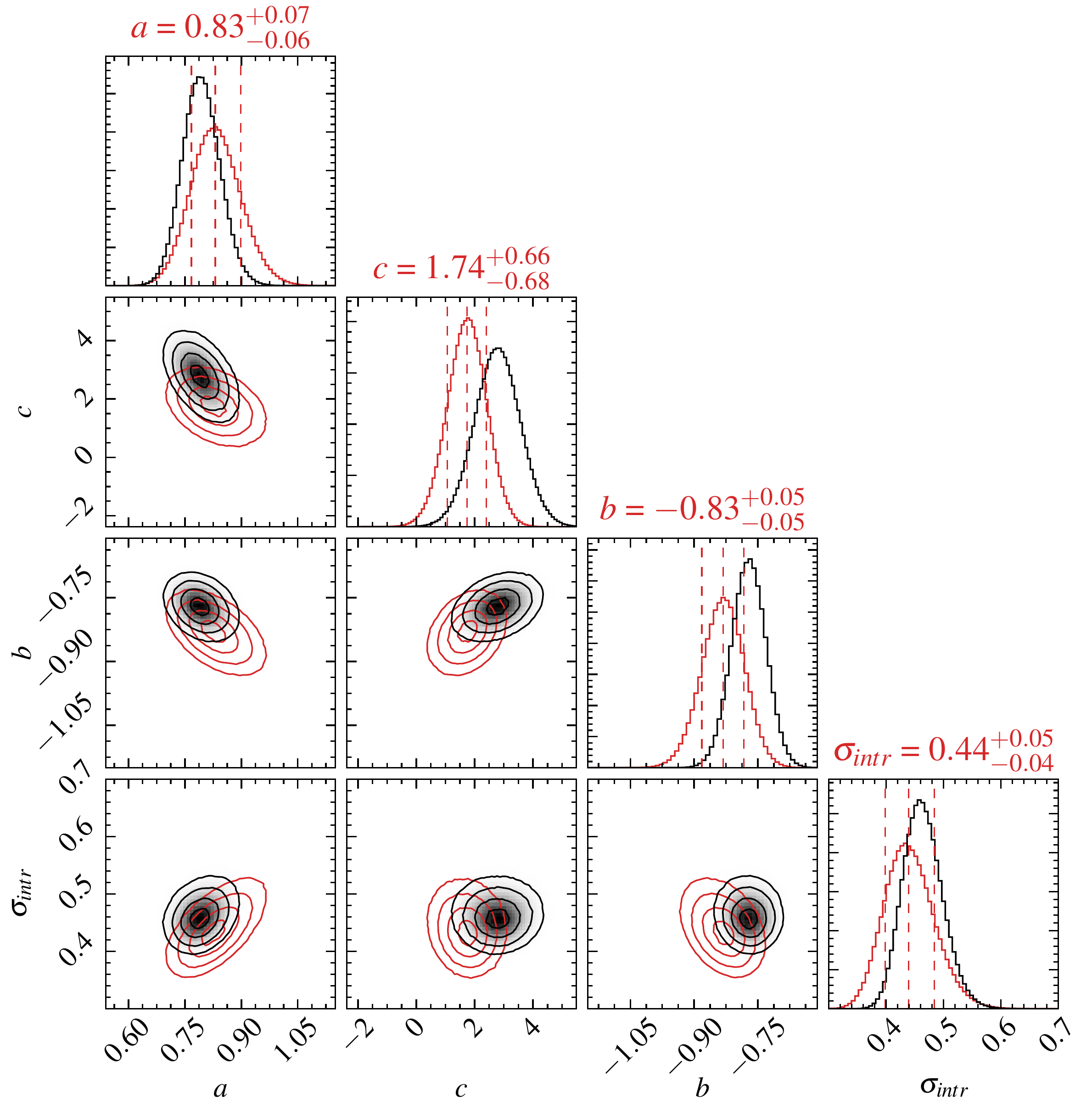}
  \caption{\label{fig:triangle} Projections of the 4D posterior distribution for
    the model parameters: slope ($a$), evolution ($c$), normalisation ($b$) and
    intrinsic scatter ($\sigma_{\mathrm{intr}}$).  The histograms on top show
    the marginalised distributions of the model parameters.  The bias-corrected
    posterior median value and the $16^{\text{th}}$ and $84^{\text{th}}$
    percentile are denoted by the dashed lines and by the values above the
    histograms.  The contours show the 0.5, 1, 1.5 and 2 $\sigma$ levels.  The
    posterior directly from the modelling is shown in black, red indicates the
    posterior after applying the bias correction (\Eq{eq:inv-trans}). Figure
    created using the \textsf{corner.py} module
    \protect\citep{Foreman-Mackey2016}.}
\end{figure*}

\begin{figure}
  \centering
  \includegraphics[width=\columnwidth]{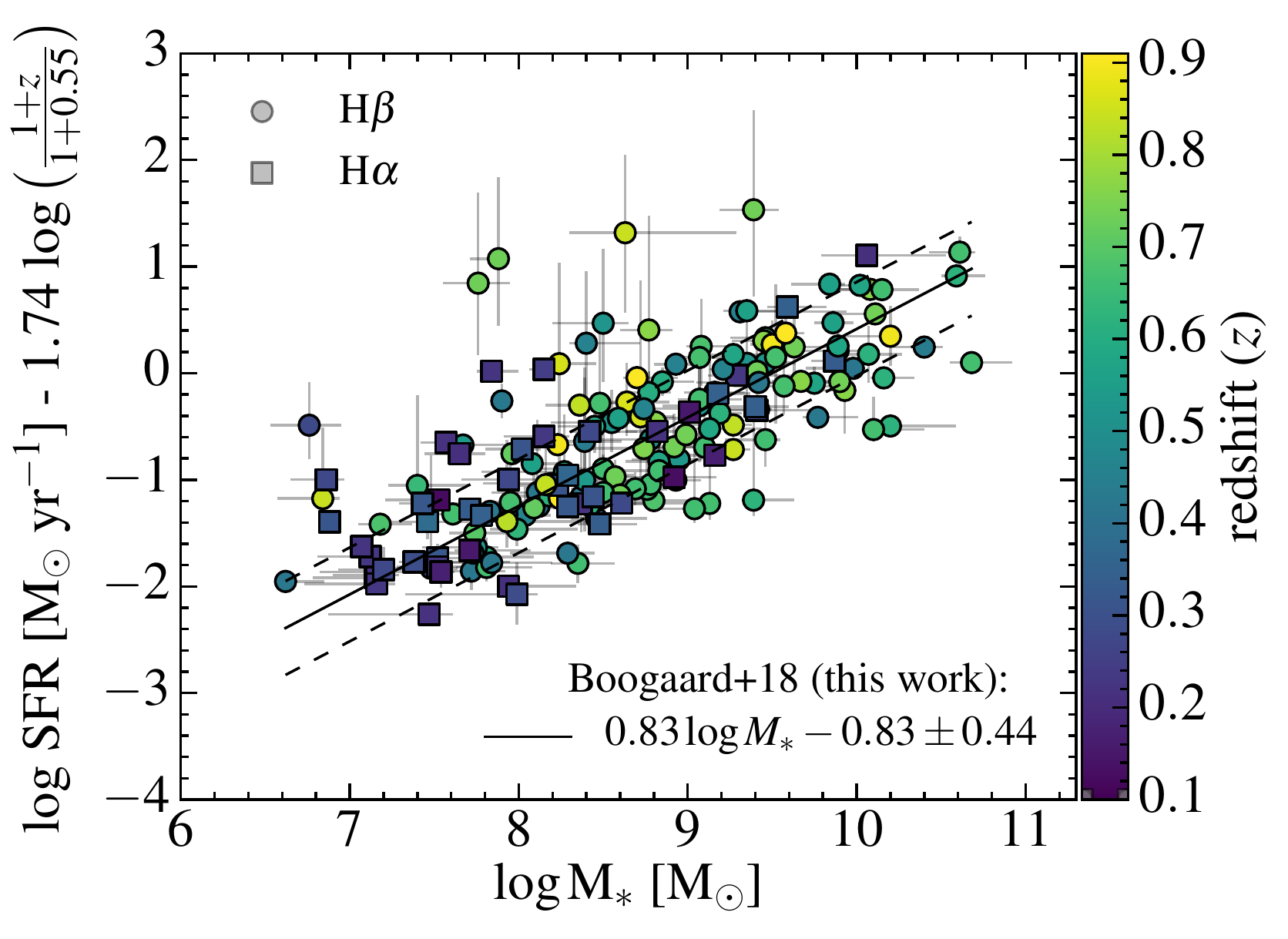}
  \caption{\label{fig:m*-z-sfr} The best-fit star formation sequence for the
    \samplesize\ star-forming galaxies observed with MUSE.  The symbols indicate
    the dust-corrected tracer used to infer the SFR.  The solid line shows
    best-fit relation, as presented in \Eq{eq:best-fit-corr}, and the dashed
    lines show the $1\sigma$ intrinsic scatter.  We subtract the evolution from
    the y-axis and scale to the average redshift of the sample; $z=0.55$.  After
    accounting for evolution, the galaxies clearly follow the star formation
    sequence, down to the lowest masses and SFRs.  The slightly larger fraction
    of galaxies that scatter into the high-mass, low-SFR regime may be a result
    of the flattening of the relation above $M_{*} = \SI{e10}{\solarmass}$.}
\end{figure}

\subsection{Global sample}
\label{sec:global-sample}

With a reliable SFR estimate in hand, we can turn to the star formation sequence
between $\samplezmin < z < \samplezmax$ as observed by MUSE.  \Fig{fig:m*-sfr}
shows a plot of stellar mass ($M_{*}$) versus star formation rate (SFR) for all
the galaxies in the sample.  The figure is based on two dust-corrected SFR
indicators: the \Hbeta\ and \Halpha\ luminosities (Eqs. \eqref{eq:sfr-hb} and
\eqref{eq:sfr-ha}).  The vertical grey lines indicate the errors in
($\log M_{*}$, $\log\mathrm{SFR}$) for each of the individual galaxies.  The
mean average error on the SFR is $\approx 0.2$ dex in both the HUDF and the
HDFS.

We are able to detect star formation in galaxies down to star formation rates as
low as \SI{0.003}{\solarmass\per\year}.  The galaxies appear to follow the
$M_{*}$-SFR trend closely over the complete mass range, down to the lowest
masses we can probe here $\sim \SI{e7}{\solarmass}$.  At the high-mass end it
appears we are starting to witness a flattening off of the trend, although we
are primarily sensitive to the intermediate and low-mass galaxies.

We model the $M_{*}$-SFR relation with the Bayesian MCMC methodology described
in detail in \Sec{sec:modelling}.  We show the resulting posterior probability
density distribution for the parameters in \Fig{fig:triangle}.  By marginalising
over the various parameters, we recover the posterior probability distributions
for the individual parameters of interest ($a,c,b,\sigma_{\text{intr}}$).  These
are plotted as histograms above the various axes in \Fig{fig:triangle}.  By
taking the median and the $16^{\text{th}}$ and $84^{\text{th}}$ percentile from
the posterior distributions we derive the median posterior value and a
$1 \sigma$ confidence interval for the parameters of interest.

The (uncorrected) best-fit (i.e. median posterior) parameters of the
distribution (with their confidence intervals) that describe the star formation
sequence are:
\begin{align}
  \label{eq:best-fit}
  \log \mathrm{SFR}[\si{\solarmass\per\year}] =\; & \slope \log\left(\frac{M_*}{M_{0}}\right) \intercpt \nonumber\\
                                                &+ \zevol \log\left(\frac{1+z}{1+z_{0}}\right) \pm \sig, %
\end{align}
analogous to \Eq{eq:linear}.  The final term represents the intrinsic scatter
(\sigtext) in the vertical ($\log \text{SFR}$) direction.  We note that while it
is a perfectly valid option for the parameterisation of the likelihood, the
posterior distribution does not favour models with zero intrinsic scatter.

\Fig{fig:triangle} shows that some correlations exist between the different
parameters of the model, which is expected.  The strongest correlation exists
between slope and redshift evolution as a less steep slope requires more
evolution in the normalisation to be compatible with the data.  The complete
covariance matrix between the different parameters is:
\begin{align}
  \label{eq:postcov}
  \Sigma(a, c, b,\sigma_{\mathrm{intr}}) &=
  \begin{pmatrix}
    \covmatrix
  \end{pmatrix}.
\end{align}

We correct the posterior for observational bias, by applying \Eq{eq:inv-trans},
which is indicated by the red contours in \Fig{fig:triangle}.  This yields a
steeper slope, with a significantly shallower redshift evolution:
\begin{align}
  \label{eq:best-fit-corr}
  \log \mathrm{SFR}[\si{\solarmass\per\year}] =\; & \corrslope \log\left(\frac{M_*}{M_{0}}\right) \corrintercpt \nonumber\\
                                                &+ \corrzevol \log\left(\frac{1+z}{1+z_{0}}\right) \pm \corrsig, %
\end{align}
At the same time, the transformation has little effect on the intrinsic scatter.
The covariance in the corrected posterior is essentially the same as the
uncorrected one, with a slight increase in covariance with intrinsic scatter.
\begin{align}
  \label{eq:postcov-corr}
  \Sigma(a, c, b,\sigma_{\mathrm{intr}}) &=
  \begin{pmatrix}
    \corrcovmatrix
  \end{pmatrix}.
\end{align}

We compare the generative distribution (i.e. \Eq{eq:best-fit}) with the data in
\Fig{fig:m*-z-sfr}.  As the plane is three dimensional, we show a projection
where we have subtracted the evolution with redshift from the y-axis.  Overall,
the distribution appears to describe the data very well and the scatter in the
observations has tightened with respect to \Fig{fig:m*-sfr}.  For a more
familiar representation we also show the resulting star formation sequence in
the right panel of \Fig{fig:m*-sfr}, for a number of different redshifts.

\subsection{Low-mass sample $(\log M_{*} [M_{\odot}] < 9.5)$}
\label{sec:low-mass-sample}
We are primarily interested in the low-mass end of the star formation sequence.  Our deep
MUSE sample spans a significant mass range, between
$\log M_{*}[\si{\solarmass}] = 6.5 - 11$.  As several studies have suggested
different characteristics for the star formation sequence above and below a turnover mass
of $M_{*} \sim 10^{10} \si{\solarmass}$ \citep[e.g.][]{Whitaker2014, Lee2015,
  Schreiber2015}, we repeat the above analysis excluding galaxies above a
certain mass threshold.  To be on the conservative side, we choose this mass
threshold to lie at $M_{*} = 10^{9.5} \si{\solarmass}$.  This excludes
$\lowmsampleexcl/\samplesize \approx 17.5\%$ of the sample.  We include this
threshold as a dashed vertical line in \Fig{fig:m*-sfr}.  We then repeat the
modelling identically to what has been described in the previous sections.

The bias-corrected star formation sequence for galaxies that have a stellar mass
below $M_{*} < 10^{9.5} \si{\solarmass}$ is:
\begin{align}
  \label{eq:best-fit-lowmass}
  \log \mathrm{SFR}[\si{\solarmass\per\year}] =\; &\corrlowmslope \log\left(\frac{M_*}{M_{0}}\right) \corrlowmintercpt \nonumber\\
                                                &+ \corrlowmzevol \log\left(\frac{1+z}{1+z_{0}}\right) \pm \corrlowmsig. %
\end{align}
The result is essentially the same, with the main difference being a steeper
redshift evolution.  All parameters are within errors consistent with the
relation for our complete sample (also for the uncorrected values, see
\autoref{tab:parameters}).  This reflects the fact that we are primarily
sensitive to the low-mass end of the galaxy sequence.  As this fit utilises only
a part of the data we will refer primarily to the fit based on all the data,
\Eq{eq:best-fit-corr}, as the main result in the remainder of the paper.  We
report the (un)corrected values for all the fits in \autoref{tab:parameters}.

\subsection{The effect of redshift bins (2D)}
\label{sec:2D-relations}
Most previous studies have not modelled the redshift evolution of the star
formation sequence directly, but have instead divided the data into redshift
bins and adopted a non-evolving relation:
$\log \mathrm{SFR} = a \log M_{*} + b$.  To facilitate the comparison with the
literature, we adapt our model to fit the relation in the
$(\log M_{*}, \log \mathrm{SFR})$-plane, without taking the redshift evolution
into account.  This is easily done, by taking a two-dimensional version of our
likelihood, disregarding the second, $\log(1+z)$-component in
Eq. (\ref{eq:mean})--(\ref{eq:prior}) --- the rest of the modelling is be
identical.  We note that we still take both heteroscedastic errors as well as
intrinsic scatter into account (see \Sec{sec:definition}), however, we do not
apply the bias correction.

We model both the entire redshift range, as well as the $0.1 < z < 0.5$ and
$0.5 < z < 1.0$ range separately (similar to other studies).  The results are
collected in \autoref{tab:parameters}.  For the full sample the slope is
significantly steeper than when we take into account the redshift evolution,
when comparing to our uncorrected fits:
\begin{equation}
  \label{eq:best-fit-twod}
  \log \mathrm{SFR}[\si{\solarmass\per\year}]= \twodslope \log\left(\frac{M_*}{M_{0}}\right)  \twodintercpt.
\end{equation}

This is also the case for the smaller samples in both redshift bins, although
the results are consistent with \Eq{eq:best-fit} within the error bars (which
are larger due to lower number statistics).  The resulting relations are
\begin{align}
  \label{eq:best-fit-twod-0-0p5}
  \log \mathrm{SFR}[\si{\solarmass\per\year}] = \lowzslope \log\left(\frac{M_*}{M_{0}}\right) \lowzintercpt,
\end{align}
for $0.1 < z \leq 0.5$ and
\begin{align}
  \label{eq:best-fit-twod-0p5-1}
\log \mathrm{SFR}[\si{\solarmass\per\year}] = \highzslope \log\left(\frac{M_*}{M_{0}}\right) \highzintercpt
\end{align}
for $0.5 < z < 1.0$.

Given the significant evolution we found in the star formation sequence with
redshift, this result is expected.  While incidently these slopes are similar to
our corrected fits, we caution that this does not imply that not modelling the
redshift evolution can circumvent biases introduced by flux-limited
observations.

\begin{table*}
  \centering
  \caption{\label{tab:parameters} Star formation sequence parameters for
    different samples (for a full description of the different samples, see
    \Sec{sec:results}).}
  \begin{tabular}{lccccc}
    \toprule
    Sample                              & Size        & $a$         & $b$             & $c$          & $\sigma_{\mathrm{intr}}$\\
    \midrule
    3D                                  & \multicolumn{5}{c}{$\log{\mathrm{SFR}[\si{\solarmass\per\year}]} = a \log{(M_{*}/M_{0})} + b + c\log{(1+z)/(1+z_{0})}$} \\
    \midrule
    Full                                & \samplesize     &  \slope     &  \intercpt      &  \zevol      & \sig \\
    $\log M_{*}[\si{\solarmass}] < 9.5$ & \lowmsamplesize &  \lowmslope & \lowmintercpt   &  \lowmzevol  & \lowmsig \\
    \midrule
    \multicolumn{6}{l}{3D -- bias corrected (via \Eq{eq:inv-trans})} \\
    \midrule
    Full                     & \samplesize     &  \corrslope     &  \corrintercpt      &  \corrzevol      & \corrsig \\
    $\log M_{*}[\si{\solarmass}] < 9.5$  & \lowmsamplesize &  \corrlowmslope & \corrlowmintercpt   &  \corrlowmzevol  & \corrlowmsig \\
    \midrule
    2D                                  & \multicolumn{5}{c}{$\log{\mathrm{SFR}[\si{\solarmass\per\year}]} = a \log{(M_{*}/M_{0})} + b$} \\
    \midrule
    Full                                & \samplesize     &  \twodslope &  \twodintercpt  &  & \twodsig \\
    $ 0.1 < z \leq 0.5$                        & \lowzsample     & \lowzslope  &  \lowzintercpt  &  & \lowzsig \\
    $ 0.5 < z < 1.0$                           & \highzsample    & \highzslope &  \highzintercpt &  & \highzsig \\
    \bottomrule
  \end{tabular}
  \tablefoot{$M_0 = \SI{e8.5}{\solarmass}$ and $z_{0} = 0.55$.}
\end{table*}

\section{Discussion}
\label{sec:discussion}

We have modelled the star formation sequence down to \SI{e8}{\solarmass} at
$\samplezmin < z < \samplezmax$ using a Bayesian framework (\Sec{sec:modelling})
that takes into account both the heteroscedastic errors on the observations as
well as the intrinsic scatter in the relation.  One major advantage of our
framework is that we simultaneously model both the slope and the evolution in
the $M_{*}$-SFR relation, while most previous studies have modelled these
separately by dividing their sample into different redshift bins.  As
demonstrated in \Sec{sec:2D-relations}, these results are not necessarily
consistent, which can be attributed to evolution taking place within a single
redshift bin.  Another important difference is that we use the Balmer lines to
trace the (dust-corrected) star formation, while most other recent studies have
relied on SFRs derived from UV+IR/SED-fitting, using different dust corrections
\citep{Whitaker2014,Lee2015,Schreiber2015,Kurczynski2016}.

As described in \Sec{sec:global-sample}, we have found that the star formation
sequence (shown in \Fig{fig:m*-sfr} and \Fig{fig:m*-z-sfr}) is well described by
\Eq{eq:best-fit-corr} (see also \autoref{tab:parameters}). We now compare our
results to other literature measurements and discuss each aspect of the star
formation sequence separately, i.e. the redshift evolution, intrinsic scatter
and the slope.  We focus particularly on the slope, for which we find the
strongest constraints, and continue with a discussion of the physical
implications of our results.

\subsection{Comparison with the literature}
\label{sec:low-mass-end}
\subsubsection{Evolution with redshift}
\label{sec:evolution-with-redshift}
We find that the normalisation in the star formation sequence increases with
redshift as $(1+z)^{c}$ with \corrzevoltext\ (\corrlowmzevol\ for
$M_{*} < \SI{e9.5}{\solarmass}$).  The fact that the normalisation of the star
formation sequence increases with redshift is well known and attributed to the
change in cosmic gas accretion rates and gas depletion timescales.  Most studies
have probed the higher mass regime and report values in the range of
$\mathrm{sSFR} \equiv \mathrm{SFR}/M_{*} \propto (1+z)^{2.5-3.5}$ at $0 < z < 3$
\citep[e.g.][]{Oliver2010, Karim2011, Ilbert2015, Schreiber2015, Tasca2015}.
Looking specifically at the low-mass regime, \cite{Whitaker2014} reports
$\text{sSFR} \propto (1+z)^{1.9}$, similar to our result.  Their more massive
end indeed shows stronger evolution $\text{sSFR} \propto (1+z)^{2.2 - 3.5}$.
\cite{Lee2015} on the other hand, find much steeper evolution, with
$\mathrm{sSFR} \propto (1+z)^{4.12 \pm 0.1}$.  We note that our parameterisation
assumes a power-law type of evolution of the star formation sequence with
redshift.  We have decided to stick to this very common first-order
approximation.  Still, one should keep in mind that a more complex evolution
with redshift is possible, both non-linear in time as well as a different
evolution in different mass regimes.  We do not find strong constraints on the
redshift evolution due to our relatively small redshift range from $z=0.1$ to
$z=0.91$.  Still, the results from \Sec{sec:2D-relations} show that it is
important to take the redshift evolution into account, in order to get a robust
constraint on the slope.

\subsubsection{Intrinsic scatter}
\label{sec:intrinsic-scatter}
Constraining the intrinsic scatter in the star formation sequence has proven to
be challenging as one has to separate the intrinsic scatter from the
measurement error \citep[e.g.][]{Noeske2007a, Salim2007, Salmi2012,
  Whitaker2012, Guo2013, Speagle2014, Schreiber2015}.  This challenge in
particular motivates our adopted model, which directly constrains the amount of
intrinsic scatter in the relationship, even in the presence of measurement
errors.  Meanwhile, our measurements are not affected by binning, e.g. we do
not boost the scatter artificially because of evolution of the star formation sequence
within a single bin.

In our best fit model we find \corrsigtext\ dex, which is larger than the value
of $\sim 0.2 - 0.4$ dex that is commonly found
\citep[e.g.][]{Speagle2014,Schreiber2015}.  \cite{Kurczynski2016} determined an
intrinsic scatter of $\sigma_{\rm intr} = 0.427 \pm 0.011$ in their lowest
redshift bin ($0.5 < z < 1.0$) in the HUDF, similar to our result, but found
significantly smaller scatter at higher redshifts.  They determined the
intrinsic scatter by decomposing the total scatter ($\sigma_{\rm Tot} = 0.525$)
using the covariance matrix between $M_{*}$ and SFR determined from their SED
fitting.

There are several effects that could potentially affect the scatter.
Measurement outliers are not a cause of concern for the intrinsic scatter as
they are taken into account by the likelihood approach.  However, if galaxies
are included in the sample that are not on the $M_{*}$-SFR relation, such as
red-sequence galaxies or starbursts, then these might artificially increase the
scatter.  We argue that the former is unlikely as our selection criteria based
on the 4000 \AA\ break and the \Halpha\ or \Hbeta\ equivalent width effectively
remove all red-sequence galaxies from the sample.  On the other hand, our sample
does include a small number of galaxies that are offset from the relation
towards high SFRs.  We verified however that removing all galaxies with a
$\mathrm{sSFR} > \SI{10}{Gyr^{-1}}$ from the sample does not significantly
increase or decrease the scatter.

Hypothetically, if the error bars on the SFR are underestimated, this will
artificially boost the intrinsic scatter in the relationship.  To determine the
influence of the size of the error bars we redid the modelling while folding in
an additional error on the SFR of 0.2 dex in quadrature (effectively doubling
the average error bars); this decreased the scatter by 20\% to $\sim 0.4$~dex.
The sample size does not seem to affect the measurement and splitting our sample
did not yield significantly larger scatter (see \Sec{sec:2D-relations}).

Assuming our measured scatter is real, it might be that previous studies have
underestimated the amount of intrinsic scatter.  One potential danger might lie
in the derivation of both stellar mass and SFR from the same photometry.
Especially in SED modelling this might introduce correlations between $M_{*}$ and
SFR as both are regularised through the same star formation history in the model
spectrum which could artificially decrease the scatter.

More physically, the difference could also in part be due to the fact that the
Balmer lines trace the SFR on shorter timescales (stars with ages \SI{\leq
  10}{Myr} and masses \SI{>10}{\solarmass}) than the UV does (ages of \SI{\leq
  100}{Myr} and masses \SI{>5}{\solarmass}; e.g. \citealt{Kennicutt1998,
  Kennicutt2012}).  Simulations have indeed found that SFRs averaged over
timescales decreasing from $10^{8}$ to $10^{6}$ yr could be significantly larger
\citep{Hopkins2014, Sparre2015}, particularly if star formation histories are
bursty \citep[e.g.][]{Dominguez2015, Sparre2017}.

Furthermore, as the recent star formation histories of low-mass galaxies are
more diverse, it can be expected that there is more scatter in the star
formation sequence at low stellar masses.  This indeed has been predicted by
simulations \citep[e.g.][]{Hopkins2014, Sparre2017} as well as semi-analytical
models \cite[e.g.][]{MitraS_17a}.  Observing such a trend requires a large and
highly complete sample of galaxies over an extended mass range and hence
evidence has been inconclusive.  Using a large sample of galaxies from the SDSS,
\cite{Salim2007} reported a decrease in the scatter of $-0.11~\mathrm{dex}^{-1}$
from \SI{e8}-\SI{e10.5}{\solarmass}, but such a trend with mass has not been
confirmed by studies at higher masses \citep{Whitaker2012, Guo2013,
  Schreiber2015, Kurczynski2016}.  Recently though, \cite{Santini2017} have
found indications of decreasing scatter with mass in the Frontier Fields, albeit
at higher redshifts ($z>1.3$).

A large and complete sample of galaxies, covering the ($\log M_{*}$, log SFR,
$\log(1+z)$)-space, with independent stellar mass and SFR estimates, is required
to get a firm handle on the intrinsic scatter in the star formation sequence.

\subsubsection{Slope}
\label{sec:slope}
\begin{figure*}
  \centering
  \includegraphics[width=\textwidth]{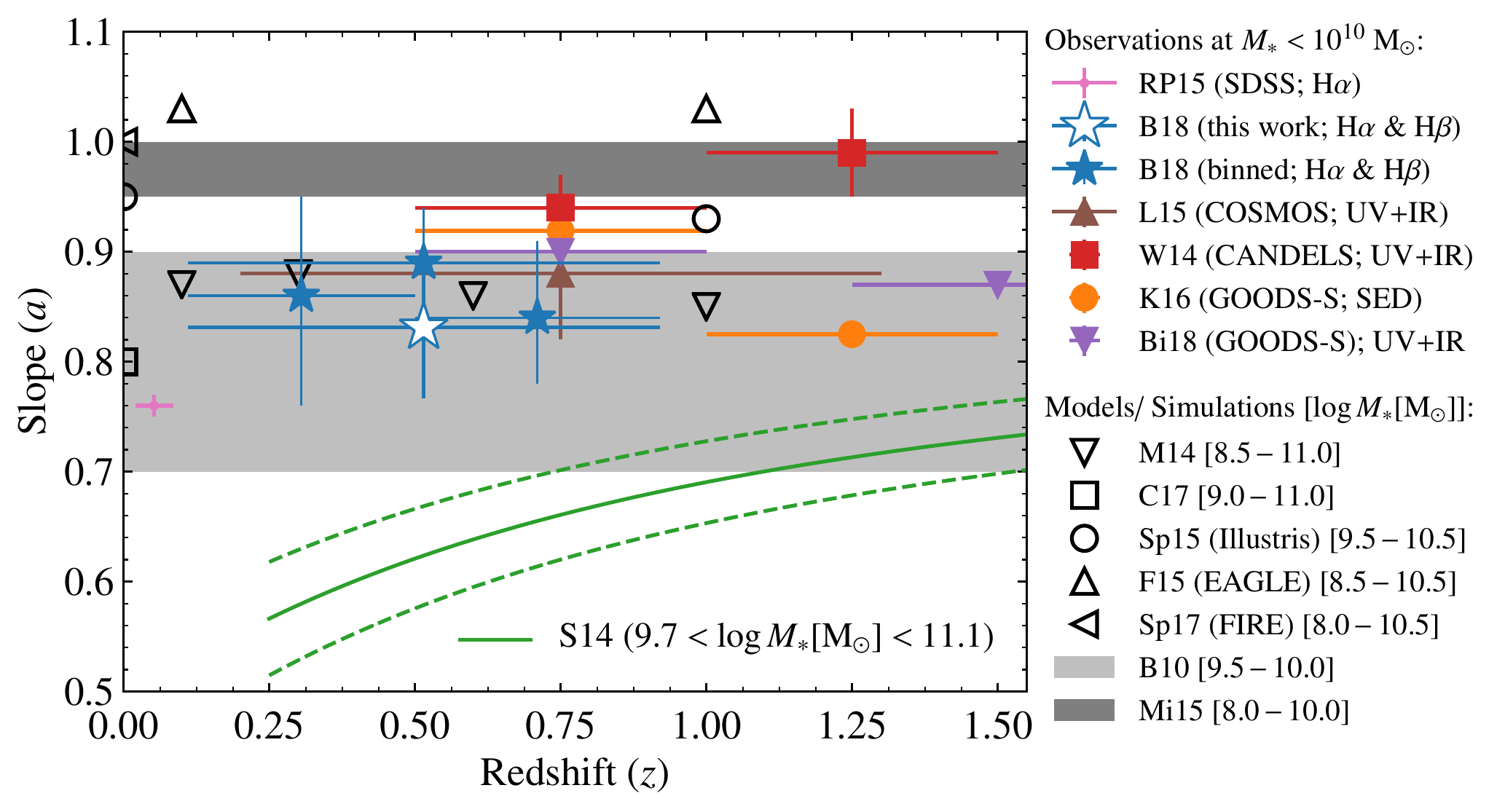}
  \caption{(a) \label{fig:params} A comparison of the slope ($a$) as a function
    of redshift ($z$) with studies from the literature that extend down to
    $M_{*} < \SI{e10}{\solarmass}$ near our redshift range.  Our best-fit
    (bias-corrected) slope from \Eq{eq:best-fit-corr} is shown by the large
    star.  As most studies have probed the slope in bins of redshift, we also
    include our results obtained using non-evolving redshift bins
    (Eqs. \eqref{eq:best-fit-twod}, \eqref{eq:best-fit-twod-0-0p5} and
    \eqref{eq:best-fit-twod-0p5-1}; smaller blue stars).  The literature results
    are from \protect\citet[][RP15]{Renzini2015},
    \protect\citet[][K16]{Kurczynski2016}, \protect\citet[][B18]{Bisigello2017}
    and the low-mass ($M_{*} \lesssim \SI{e10}{\solarmass}$) power-law slopes
    from \protect\citet[][W14]{Whitaker2014} and \protect\citet[][L15]{Lee2015}.
    We also add \protect\citet[][SP14]{Speagle2014} for reference, though it is
    inferred at higher masses.  We indicate the field and SFR-tracer in
    brackets, though note that distinct calibrations for the same tracer may be
    used in different studies.  In addition, we add the slopes predicted by
    (semi-)analytical models; \protect\citet[][B10]{BoucheN_10a},
    \protect\citet[][M14]{Mitchell2014}, \protect\citet[][Mi15]{MitraS_15a},
    \protect\citet[][C17]{Cattaneo2017}, and hydrodynamical simulations;
    \protect\citet[][Sp15]{Sparre2015}, \protect\citet[][F15]{Furlong2015},
    \protect\citet[][Sp17]{Sparre2017}.}
\end{figure*}

We find a best-fit (median posterior) slope of the star formation sequence of
\corrslopetext\ (log SFR $\propto a \log M_{*}$).  This slope is determined from
galaxies that are more than an order of magnitude lower in mass than most
earlier studies at $z>0$, i.e. at \SI{e8} to \SI{e10}{\solarmass}, whereas most
previous studies \citep[e.g.][]{Speagle2014, Lee2015, Schreiber2015} have been
primarily sensitive to a higher mass range from \SI{e9.5}{\solarmass} to
\SI{e11}{\solarmass}.  For reference, we plot the polynomial fit from
\cite{Whitaker2014} (down to their mass completeness limit, based on stacking)
in \Fig{fig:m*-sfr}.

Recent studies have typically observed a shallower slope at the high-mass end,
i.e. above \SI{e10}{\solarmass} \cite[e.g.][]{Whitaker2014}.
\cite{Gavazzi2015} find a turnover mass of $M_{*} \sim \SI{e9.7}{\solarmass}$ at
$z=0.55$ (after converting their result to a Chabrier IMF), increasing with
redshift.  As discussed in \Sec{sec:low-mass-sample}, excluding galaxies above
$M_{*} > \SI{e9.5}{\solarmass}$ has no significant effect on the slope.  Only
$15/\samplesize \approx 8.5\%$ of galaxies in our sample have
$M_{*} > \SI{e10}{\solarmass}$ and thus our result is not very sensitive to this
turn-over.  In light of this, we limit the following discussion to studies which
specifically probe the mass range below the turnover of the star formation
sequence.

Our best-fit slope of \corrslope\ is compared to the values found by other
recent studies in \Fig{fig:params} where we focus on studies with similar
redshift ranges (i.e. $0<z<1$) and which extend well below
$M_{*} <\SI{e10}{\solarmass}$.  The slope in this regime is notably steeper than
the consensus relation from \cite{Speagle2014} who reported $a = 0.6 - 0.7$ at
our redshifts, due to the fact that this compilation is for a mass range of
$\log M_{*}[\si{\solarmass}] = 9.7 - 11.1$, where the slope is significantly
shallower.  Our slope is shallower than the low-mass power-law slope from
\cite{Whitaker2014} ($a = 0.94 \pm 0.03$ for $M_{*} < \SI{e10.2}{\solarmass}$)
from the 3D-HST catalogues in CANDELS, but is consistent with the global slope
of $a = 0.88 \pm 0.06$ reported by \cite{Lee2015} in a large sample of
star-forming galaxies in COSMOS.  \cite{Kurczynski2016} have also presented a
characterisation of the star formation sequence in the HUDF, based on the
CANDELS/GOODS-S \citep{Santini2015} and UVUDF \citep{Rafelski2015} catalogues.
In their lowest redshift bin ($0.5 < z < 1.0$), which goes down to
$M_{*} \sim 10^{7.5} \si{\solarmass}$ they find a slope of $a=0.919 \pm 0.017$,
which is also steeper (marginally consistent) compared to what find.  We note
that they determined both masses and SFRs from the SED modelling, taking into
account the correlations between the parameters, as their study was focused
particularly on measuring the intrinsic scatter, see
\Sec{sec:intrinsic-scatter}.  In the same field \cite{Bisigello2017} find a
slope of $0.9 \pm 0.01$ ($0.5 \leq z < 1.0$), after selecting galaxies with
$\log \mathrm{sSFR[\si{\per\giga\year}]} < - 9.8$.

The Sloan Digital Sky Survey (SDSS; \citealt{York2000,Abazajian2009}) serves as
a natural reference for Balmer line-derived SFRs in the local universe and since
\cite{Brinchmann2004} different studies have derived the star formation sequence
slope \citep[e.g,][]{Salim2007, Elbaz2007}.  The most recent of these is
\cite{Renzini2015}, who measure the slope of the ridge line in the
$M_{*}-N\times \mathrm{SFR}$-plane (where $N$ is the number of galaxies in every
$M_{*}$-SFR bin) and find $a=0.76\pm0.01$, which is significantly flatter than
our results.

Taken at face value, our slope of \corrslopetext\ is inconsistent with a linear
slope ($a=1$).  A value (close to) unity may have been expected on the basis of
simulations (see next section), which is also evident from the fact that several
parameterisations of the star formation sequence asymptote to a linear relation
at low mass \citep[e.g.][]{Schreiber2015, Tomczak2016}.  An independent
motivation for a near-linear value comes from the fact that there is very little
evolution in the faint slope of the stellar mass function of star-forming
galaxies up to $z=2$ (see, e.g. \cite{Tomczak2014, Davidzon2017} for recent
results).  To first order, this may implies self-similar mass growth for
low-mass galaxies (i.e. constant sSFR which implies a linear slope for the star
formation sequence), unless balanced by mergers \citep{Peng2014}.
\cite{Leja2015} investigated the link between the slope of the star formation
sequence and the stellar mass function.  While they do not provide precise
constraints on the low-mass slope at low redshift (due to the challenge of
disentangling growth through star formation and mergers), their results indicate
that a sub-linear low-mass slope is still consistent with the stellar mass
functions at $z<1$.

\subsubsection{Evolution of the low-mass slope}
\label{sec:evolution-low-mass}

Combining results from the local universe out to redshift $z \sim 6$,
\cite{Speagle2014} found evidence for an evolving slope at the high-mass end
($M_{*} > \SI{e9.7}{\solarmass}$), where the slope gets shallower with redshift
\cite[cf.][Fig. 5]{Abramson2016}.  Given the turnover in the star formation
sequence at high mass, it is important to disentangle to what extent the
evolution in the slope is due to different studies being sensitive to distinct
mass regimes.  Our data are too sparse in redshift space to simultaneously
constrain the evolution of the slope (and hence we have adopted a single
power-law slope for the sequence).

In light of the potential redshift evolution of the slope, we plot the slope as
a function of redshift in \Fig{fig:params}, compared to literature results which
probe the mass range $M* < \SI{e10}{\solarmass}$ at $z < 1.5$.  \Fig{fig:params}
provides evidence for evolution of the low-mass slope with redshift.  However,
we caution against a too strong interpretation of such a trend as the literature
suffers from studies probing distinct mass ranges (sometimes including the
turn-over regime).  What further complicates a fair comparison is that different
tracers of star formation probe different timescales and additionally use
varying dust corrections, which are not necessarily consistent
\citep[e.g.][]{Davies2016}.  A consistent analysis of the low-mass galaxy
population out to higher redshifts is important to quantify potential evolution
in the low-mass slope.

\subsection{The MS slope --- a quantitative comparison to models}
\label{sec:model}

The galaxy main sequence (MS) is a natural outcome of hydrodynamical models
(e.g. Fig. 1b in \citealt{Bouche2005}; \citealt{Dave2008, GenelS_14a,
  Torrey2014, Kannan2014, Hopkins2014, Sparre2015, Furlong2015}) and in
semi-analytical models \citep[e.g.][]{Somerville2008, DuttonA_10a,
  Cattaneo2011, Mitchell2014, Henriques2015, Hirschmann2016, Cattaneo2017}.
These models have reported a slope (and scatter) that, in general, is broadly
consistent with observations, but the quantitative details regarding the slope
and/or the evolution of the main sequence often do not match observations.

Since the pioneering work of \citet{Daddi2007} and \citet{Elbaz2007}, it has
been noted that the redshift evolution of the main sequence normalisation, in
particular around $z=2$, is a challenge for models \citep[e.g.][]{Dave2008,
  Damen2009, BoucheN_10a, DuttonA_10a, DekelA_14a, Torrey2014, GenelS_14a,
  Mitchell2014, Furlong2015, Sparre2015, Abramson2016, Santini2017}.  Here, we
focus on a quantitative comparison of the slope of the main sequence
($\mathrm{SFR} \propto M_{*}^a$) with various models, given that our study
yields the tightest constraint on this parameter (compared to the other
parameters in the model).

The Illustris simulations \citep{Vogelsberger2014,GenelS_14a,Sparre2015} produce
a main-sequence with a slope $a$ that is slightly sub-linear with
$a\lessapprox1.0$. In particular, \citet{GenelS_14a} noted that sSFR goes as
$\simeq-0.1$ with stellar mass and using the results from \citet{Sparre2015}, we
find that the main sequence in Illustris goes as
$\mathrm{SFR} \propto M_{*}^{\approx0.95}$.  The EAGLE simulations
\citep{Schaye2015, Crain2015} also allow an investigation of the main sequence
and \citet[][their Fig. 5]{Furlong2015}, showed that the sSFR is constant with
$M_{*}$ from \SI{e8} to \SI{e10}{\solarmass} at redshifts $z=0.1$, 1.0 and 2.0,
with a relatively steep decline above \SI{e10}{\solarmass}. Quantitatively,
below \SI{e10}{\solarmass}, the slope of the main sequence $a$ in
\citet{Furlong2015} is $a\approx1.04$.  The MS slope for the Illustris and EAGLE
simulations are shown in \Fig{fig:params} as the open circles and triangle
symbols, respectively.  In the FIRE simulations \citep{Hopkins2014},
\cite{Sparre2017} focused on studying the scatter in the main sequence for
different tracers of SFR and shows a slope of $a\approx0.98$ when using the FUV
(their Fig. 2).

The MS slope has also been a challenge for semi-analytical models because
different (regular) feedback prescriptions do not alter the MS slope as shown in
\citet{DuttonA_10a} and discussed in \citet{Mitchell2014} (however, it can alter
the slope in hydrodynamical simulations, e.g. \citealt{Haas2013a, Haas2013b,
  Crain2015}).  \citet{Mitchell2014} performed a detailed comparison between
predictions from the GALFORM semi-analytical models with observations and their
fiducial model produces a MS slope of $a\approx0.85$ (shown in \Fig{fig:params}
as the down-pointing triangles).  Recently, the semi-analytical model of
\citet{Cattaneo2017} using the GALICS2 code was set to reproduce the local
luminosity function and the local MS slope simultaneously.  Their MS slope is
$a\approx0.8$ (open square in \Fig{fig:params}), but we caution their use of an
extreme feedback model, where the mass loading $\eta$ is $\eta\propto V^{-6}$,
where $V$ is the halo virial velocity.  Such a steep scaling between galaxy mass
and wind loading is not supported by the data \citep[e.g.][]{SchroetterI_16a}.

\citet{BoucheN_10a} used a simple toy model for galaxy (self-)regulation with
which they showed that variations in feedback prescriptions or in the laws of
star formation have no impact on the MS slope. They argued that while ejective
feedback alone is not sufficient to bring the theoretical slope of the
main-sequence in agreement with observations, preventive feedback can easily do
so as several studies have shown \citep{Dave2012,Lu2015,MitraS_15a,MitraS_17a}.
However, while the MS slope of \citet{BoucheN_10a} is sub-linear with
$a\approx0.8$, a quantitative analysis reveals that the slope varies rapidly
with stellar mass, likely due to the limitations of the model. Indeed, the MS
slope of \citet{BoucheN_10a} goes from 0.7 at $M_{*}\sim\SI{e9.5}{\solarmass}$
to 0.9 at $M_{*}\sim\SI{e10.5}{\solarmass}$.  The range of values is indicated
by the light grey box in \Fig{fig:params}.

\citet{MitraS_15a} expanded the self-regulation model of \citet[and
others]{BoucheN_10a,Dave2012} with physically motivated parameters and attempted
to determine these parameters using a Bayesian MCMC approach on a set of
observed scaling relations at $0<z<2$. Their fiducial model yields a MS with a
slope that is quasi-linear with $a\sim0.95$ in our mass regime, i.e. below
$\SI{e10}{\solarmass}$. Their MS slope is shown as the dark grey band in
\Fig{fig:params}.

Generally speaking, in the low-mass regime below $\SI{e10}{\solarmass}$,
hydrodynamical simulations have steeper MS slopes with $a\approx1.0$ whereas our
estimate (\corrslopetext) at $z<1$ and recent observations covering that mass
range indicate $a<1.0$ (see \Fig{fig:params}).  The reason that models tend to
predict a steeper main sequence slope lies in the underlying feature in
hydrodynamical simulations and semi-analytical models, where the growth rate for
dark matter halos $\dot M_{\rm h}$ scales with mass as
$\dot M_{\rm h}\propto M_{\rm h}^{1.15}$ \citep{BirnboimY_07a, GenelS_08a,
  DekelA_09a, FakhouriO_08a, NeisteinE_08a}, in combination with rapid gas
cooling.

\subsection{Implications of a shallow slope}
As noted originally by \citet{Noeske2007b} and discussed in \citet{Mitchell2014}
and \citet{Abramson2016}, a MS with a sub-linear slope,
$\mathrm{SFR} \propto M_{*}^{a}$ with $a < 1$, implies downsizing where
lower-mass galaxies have longer $e$-folding time and a later onset of star
formation.  This downsizing effect would be amplified if the MS slope is
substantially flatter above \SI{e10}{\solarmass} as some studies have indicated
\citep{Whitaker2014,Schreiber2015,Lee2015,Tomczak2016}.  This turnover has
generally been attributed to either a morphological transition, such as bulge
growth \citep{Abramson2014,Lee2015,Whitaker2015}, or a reduced star formation
efficiency \citep{Schreiber2016}.

Our result, that the slope of the main sequence is sub-linear in the low-mass
regime, implies that there are processes at work which either: (1) affect the
conversion of the accreted gas into stars through increased (supernova) feedback
or a decrease in the SF efficiency; or (2) prevent the accretion of gas onto
low-mass galaxies.  These two processes might conspire with the fact that the
gravitational potential is shallower in low-mass galaxies \citep{MitraS_15a}.

In hydrodynamical simulations low-mass galaxies (up to halo masses of \SI{\sim
  e11.5}{\solarmass}) obtain their gas primarily through `cold'-accretion
\citep{Keres2005, vandeVoort2011}, where the gas is never heated to the virial
temperature, while `hot' accretion, where gas is first shock heated to the
virial temperature and then cools and accretes, is dominant for more massive
galaxies.  A candidate process is feedback from gravitational heating, due to
the formation of virial shocks \cite[e.g.][]{Faucher-Giguere2011}, which
becomes more effective at higher masses, however, can still play a role down to
halo masses of \SI{e10}{\solarmass}.  The heating of gas through winds (from
either supernovae or black hole feedback) can also prevent the gas from flowing
into the galaxy \citep{Oppenheimer2010, Faucher-Giguere2011, vandeVoort2011}, in
particular in low-mass galaxies.  However, \cite{Schaye2010} pointed out that
this type of feedback mainly has a regulatory effect on the gas infall.

As noted by \citet{DuttonA_10a}, \citet{BoucheN_10a}, and \citet{Mitchell2014},
in semi-analytical models, the MS slope is rather insensitive to the ejective
(regular) feedback mechanisms~\footnote{with mass loading $\eta\propto V^{-1}$
  or $\eta\propto V^{-2}$ for momentum or energy-driven winds, respectively.},
such as the heating of gas through winds and/or the star formation efficiency
\citep{Kennicutt1998} because they act primarily on the gas content.  Hence,
the SFR and stellar mass are affected in a similar way, leaving the slope
unchanged, unless the ejective feedback prescription is strongly mass dependent
with $\eta\propto V^{-6}$, as in \citealt{Cattaneo2017}.  In addition,
\citet{Mitchell2014} showed that the slope is also insensitive to the gas
re-incorporation prescription \citep[see also][]{MitraS_15a}.

Preventive processes \citep{BlanchardA_92a, Gnedin2000a, Mo2005, Lu2007,
  Okamoto2008} that tend to be mass dependent can more easily impact the MS
slope, the Tully-Fischer relation, and the luminosity function as argued by
\citet{BoucheN_10a}.  A preventive process which can prevent the inflow of gas
specifically in low-mass halos is photoionisation heating \citep{Quinn1996}.
While it has been argued that this process is primarily effective in dwarf
galaxies and becomes ineffective above halo masses of a few times
\SI{e9}{\solarmass} \citep[e.g.][]{Okamoto2008}, \citet{CantalupoS_10a} suggest
that photoionisation may still play a role for more massive halos if there is
significant star formation.

\section{Summary and conclusions}
\label{sec:conclusions}
We have exploited the unique capabilities of the MUSE instrument to investigate
the star formation sequence for low-mass galaxies at intermediate redshift
($\samplezmin < z < \samplezmax$).  From the large number of sources detected
with MUSE in the HUDF and HDFS we have constructed a sample of \samplesize\
star-forming galaxies down to $M_{*} \sim \SI{e8}{\solarmass}$, with a number of
objects at even lower masses (\Fig{fig:mass-histogram}).  The accurate
spectroscopic redshifts from MUSE are combined with the deep photometry
available over the HUDF and HDFS to determine a robust mass estimate for the
galaxies in our sample through stellar population synthesis modelling.

With MUSE we can detect star-forming galaxies down to SFR
$\sim \SI{e-3}{\solarmass \per \year}$ (\Fig{fig:m*-sfr}).  We show that we can
determine robust, dust-corrected SFR estimates from \Halpha\ and \Hbeta\
recombination lines, by comparing the SFRs from different tracers
(\Fig{fig:sfr-consistency}).  A dust-corrected star formation rate is inferred
from the \Halpha\ and \Hbeta\ emission lines observed with S/N > 3 in the MUSE
spectra.

We characterise the star formation sequence by a Gaussian distribution around a
plane (\Eq{eq:linear}).  This methodology is chosen to maximally exploit the
data set taking into account heteroscedastic errors.  We constrain the slope,
normalisation, intrinsic scatter, and evolution with redshift from the posterior
probability distribution via MCMC methods (\Fig{fig:triangle}).

We analyse the robustness of our model and the influence of the MUSE detection
limit on the derived properties of the star formation sequence, by determining
how well we can recover the parameters from a sample of simulated relations
(detailed in \autoref{sec:simulations}).  Using the results, we correct our
inferred parameters for observational biases.

We report a best-fit description of the low-mass end of the galaxy star
formation sequence of
$\log \mathrm{SFR} = \corrslope \log M_{*} \corrintercpt + \corrzevol \log(1+z)$
between $\samplezmin < z < \samplezmax$, shown in \Fig{fig:m*-z-sfr}.  The full
description of our parameters, including errors and normalisation, is found in
\Eq{eq:best-fit-corr}.

The intrinsic scatter around the sequence is found to be \corrsigtext\ dex (in
log SFR).  This is notably higher than the average value reported in literature
($\sim 0.3$ dex), which could be attributed to a combination of the Balmer lines
probing star formation on shorter timescales and the star formation histories of
low-mass galaxies being more diverse.

Excluding massive galaxies (with $M_{*} > \SI{e9.5}{\solarmass}$) has no
significant effect on the best-fit parameters, indicating we are primarily
sensitive to low-mass galaxies.  Notably though, we find that the slope steepens
when splitting our sample into one or multiple redshift bins, with the values
going up to
$\log \mathrm{SFR} [\si{\solarmass\per\year}] = \twodslope \log
M_{*}[\si{\solarmass}]$.  This shows the importance of taking into account the
evolution with redshift when deriving the properties of the star formation
sequence.

The slope of the star formation sequence is an important observable as it
provides information on the processes that regulate star formation in galaxies.
Our slope is shallower than most simulations and (semi-)analytical models
predict, which find a (super-)linear slope essentially due to the growth rate of
dark matter halos.  Feedback processes operating specifically in the low-mass
regime, which affect the accretion of gas onto galaxies and/or subsequent star
formation, are required to reconcile these differences.  Models suggest that
supernova feedback or a decreased star formation efficiency do not affect the
slope of the star formation sequence.  Instead, processes that prevent the
accretion of gas onto low-mass galaxies are thought to play an important role in
determining the slope of the star formation sequence in the low-mass regime.

\begin{acknowledgements}
  We would like to thank the referee for providing a constructive report that
  helped improve the quality of the paper.  LB would like to thank the
  participants of the Lorentz Center Workshop on \emph{A Decade of the
    Star-Forming Main Sequence} for beneficial discussions.  We gratefully
  acknowledge the developers of \textsc{IPython}, \textsc{Numpy},
  \textsc{Matplotlib}, and \textsc{Astropy}
  \citep{Perez2007,VanDerWalt2011,Hunter2007,Robitaille2013} and \textsc{Topcat}
  \citep{Taylor2005} for their development of the software used at various
  stages during this work.  JB acknowledges support from Funda{\c c}{\~a}o para
  a Ci{\^e}ncia e a Tecnologia (FCT) through national funds (UID/FIS/04434/2013)
  and Investigador FCT contract IF/01654/2014/CP1215/CT0003., and from FEDER
  through COMPETE2020 (POCI-01-0145-FEDER-007672).  JS acknowledges support from
  the Netherlands Organisation for Scientific Research (NWO) through VICI grant
  639.043.409.  NB and TC acknowledge funding by the ANR FOGHAR
  (ANR-13-BS05-0010-02), the OCEVU Labex (ANR-11-LABX-0060), and the A*MIDEX
  project (ANR-11-IDEX-0001-02) funded by the ``Investissements d'avenir''
  French government programme.  RB acknowledges support from the ERC advanced
  grant 339659-MUSICOS.
\end{acknowledgements}

\bibliographystyle{aa}
\bibliography{Msc1.bib}

\begin{appendix}
\section{Simulations}
\label{sec:simulations}

\subsection{Selection function and completeness}
\label{sec:selection-function-completeness}
We have selected galaxies based on the signal-to-noise of their emission lines,
without any photometric preselection.  This means the selection function is
essentially determined by the emission line sensitivity.  In general, one might
expect galaxies with higher S/N in their emission lines to have a higher SFR at
a fixed mass, or similarly, for galaxies with the same S/N to have a higher SFR
at higher redshift, which potentially introduces biases in our results.
Additionally, we can only observe galaxies that have Balmer lines in the
spectral range of MUSE ($z < 0.91$).

To investigate the influence of these selections, we determine how well we can
recover the true parameters of the star formation sequence from a set of mock
samples of galaxies, after applying the flux limit from our MUSE observations.

We determine the influence of the selection function on the inferred parameters
by simulating mock data for a range of `true' parameters.  The range of values
for each mock parameter is listed in \autoref{tab:mock-grid-input}, which
combine to form a grid of $N=\simgridsize$ points.  The extent of grid is chosen
such that it encompasses a wide range of possible parameters and we find that
the results are consistent if we enlarge the grid even further (note that, if
the grid is taken too large, non-linearities may arise at the extreme values
which potentially bias the linear transformation approach of
\Sec{sec:line-transf}).  We denote each set of parameters as
$\mathbf{x}_{\mathrm{true}, k} = \left(\hat a, \hat c, \hat b, \hat
  \sigma_{\mathrm{intr}}\right)^{T}$ with $k = 1, ..., N$.

We generate realistic mock data for each set of parameters through the following
procedure: We sample \simsamplesize\ galaxies from a uniform distribution in
both mass ($7.0 < \log M_{*}[\si{\solarmass}] < 10.5$) and redshift
($0.1 < z < 1$).  Given the mass and redshift, we compute the SFR (via
\Eq{eq:linear}), i.e. assuming a mock main sequence distribution with slope
$\hat a$ and evolution $\hat c$.  We choose our normalisation ($\hat b$) such
that a \SI{e10}{\solarmass} galaxy at $z=0$ has a SFR of \SI{1}{\solarmass/yr},
similar to our results and, e.g. the Milky Way \citep{Chomiuk2011}, i.e. we
take a zero-point of $b_{0} = -10$.  We then sample up to
$b_{\mathrm{offset}} = \pm 0.4$ dex above and below this zero-point.  We provide
each galaxy with a random offset from the main sequence (perpendicular to the
$(\log M_{*}, \log \mathrm{SFR}$)-relation) drawn from
$\mathcal{N}(0, \hat \sigma_{\mathrm{intr}})$.  Finally, we apply a random
measurement error for each galaxy in both log stellar mass and log SFR of 0.3
dex (i.e. drawn from $\mathcal{N}(0,0.3)$) and in log redshift of \num{5e-4}
dex ($\sim \mathcal{N}(0,\num{5e-4})$), similar to the observations.

We then apply the same flux limit as our shallowest MUSE observations, namely in
the \textsf{mosaic} with \SI{3e-19}{erg.s^{-1}.cm^{-2}}, and mark all `observed'
galaxies as those that fall above our detection threshold (we do not take an
additional factor for dust into account as our galaxies are not very dusty on
average).  We then fit the observed galaxies above the flux limit.  Repeating
this process \simnomcruns\ times for each individual set of parameters
$\mathbf{x}_{\mathrm{true}, }$ and marginalising over the combined posterior
distribution, we determine the corresponding recovered parameters
$\mathbf{x}_{\mathrm{out}, k} = \left( a, c, b,
  \sigma_{\mathrm{intr}}\right)^{T}$.

\begin{figure*}
  \centering
  \includegraphics[width=\textwidth]{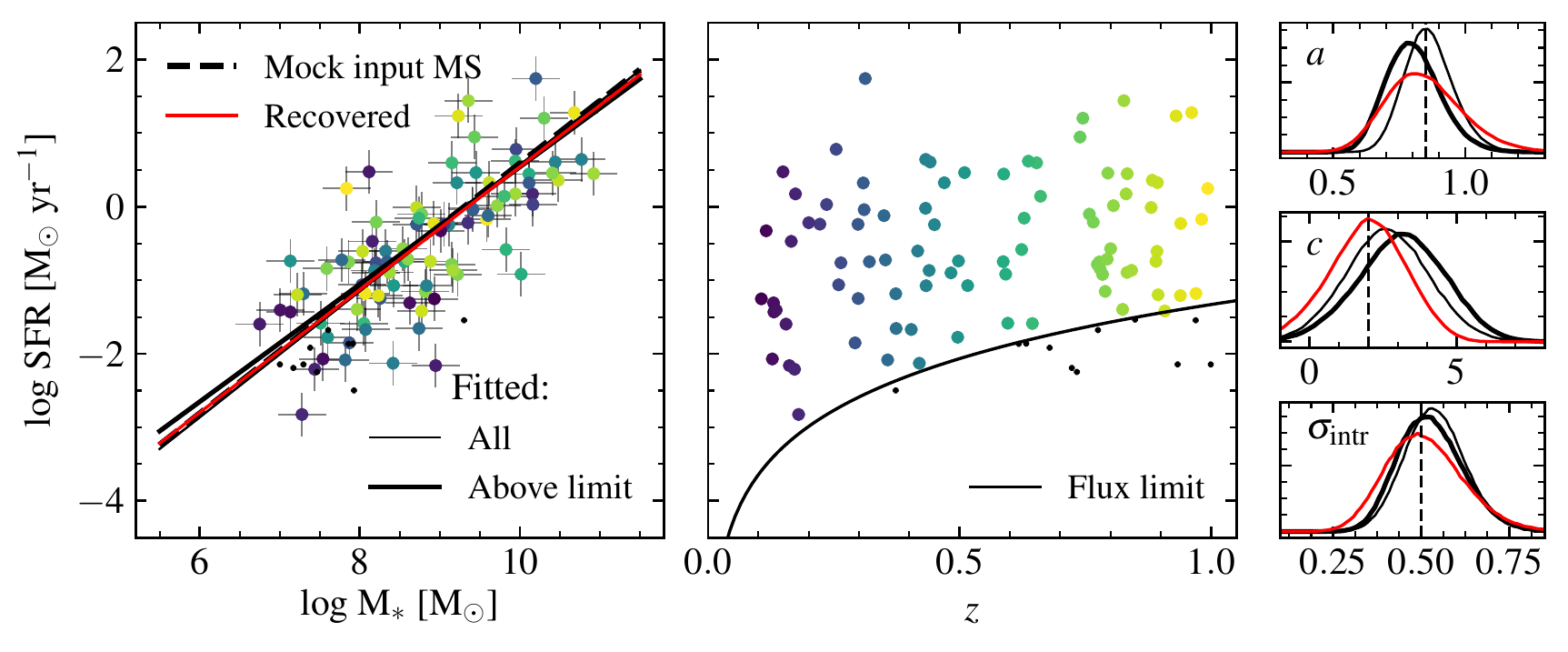}
  \caption{\label{fig:sfc} Illustration of the results of the recovery
    experiment on mock galaxies.  The points in the left and centre panels show
    one of the \simnomcruns\ realisations of \recsamplesizenocut\ galaxies in
    ($\log M_{*}$, $\log(1+z)$, $\log$ SFR)-space from a mock star formation
    sequence: $\log \mathrm{SFR} \propto a \log M_{*} + c \log(1+z)$, where in
    this particular case $a=\rexampleslopetrue$ and $c=\rexamplezevoltrue$ with
    $\sigma_{\rm intr} = \rexamplesigtrue$ dex.  The colour indicates redshift,
    unless a mock galaxy falls below the solid line in the centre panel,
    indicating the flux limit of \SI{\sim 3e-19}{erg.s^{-1}.cm^{-2}}, in which
    case it is a black point.  The rightmost panels show the marginalised
    distributions (slope, redshift evolution, and intrinsic scatter) from
    combining all \simnomcruns\ realisations.  The thin and thick black lines
    indicate the results when taking into account all mock data and only the
    data above the flux limit, respectively, and are compared to the input
    values (dashed lines).  With all data points (including noise), we can
    recover the input parameters sequence well.  When applying the flux limit a
    slight bias towards a shallower slope and steeper redshift evolution
    appears.  We plot all curves in the leftmost panel at the average redshift
    of the sample ($z_{0}$).  The red line is obtained after applying the
    correction to the fit of the data above the limit.  These recovered curves
    are plotted in the leftmost panel as well and compared to the input mock
    relation.  With our correction, we can recover the true input parameters
    well, even in the case of limited data.}
\end{figure*}

As an example, we show one the experiment for a particular set of parameters in
\Fig{fig:sfc}.  It is clear that the recovered parameters are biased towards a
shallower slope and a steeper redshift evolution.  The magnitude of this bias
depends on all the parameters and becomes more severe for steeper slopes and
shallower redshift evolutions.

To check our methods, we also fit all simulated galaxies (without discarding any
data).  Reassuringly, we recover our input parameters to within the errors, even
when simulating only 100 galaxies.  Since our actual sample size is \samplesize\
galaxies, we are in principle able to recover the true parameters of the
relation, even in the case of intrinsic scatter and heteroscedastic errors.  One
feature that does draw attention is that the redshift evolution is marginally
steeper than the input relation (but admittedly poorly constrained and still
consistent within the error).  This can be explained due to an intricacy of the
model, which assumes that the intrinsic scatter about the relation is along the
normal vector to the plane ($\sigma_{\perp}$ in \Sec{sec:definition}), i.e.
also in the $\log(1+z)$-direction.  If the data are truncated and there is a
non-zero slope $(|c| > 0)$ in redshift space, this may introduce an artificial
bias in the corresponding slope (and scatter) as the truncation boundaries are
not parallel to the normal vector.  Given the fact that our data (and mock
sample) are limited in redshift space by the spectral range of MUSE, this means
that we may have slight artificial bias towards a steeper redshift evolution.
For interpreting the intrinsic scatter this is not a problem as we can project
the scatter along the (physical) $\log \mathrm{SFR}$-axis (which is our
$\sigma_{\rm intr}$).

With our simulations in hand however, we are now in place to apply a correction
for both biases identified above.

\begin{table}
  \centering
  \caption{\label{tab:mock-grid-input} Grid values for our mock simulations.
    The normalisation $(b_{0} = -10)$ is chosen such that a
    $\SI{e10}{\solarmass}$ galaxy at $z=0$ has a SFR of
    \SI{1}{\solarmass\per\year}.}
  \begin{tabular}{lccc}
    \toprule
    & min & max & step \\
    \midrule
    $\hat a$ & 0.7 & 1.1 & 0.05 \\
    $\hat c$ & 1.5 & 4.5 & 0.5 \\
    $b_{\mathrm{offset}}$ & -0.4 & 0.4 & 0.2 \\
    $\hat \sigma_{\mathrm{intr}}$ & 0.3 & 0.6 & 0.1 \\
    \midrule
    \multicolumn{4}{l}{$\hat b = \hat a \left(\log(M_{0}) - b_{0}\right) + \hat c
    \log(1+z_{0}) + b_{\mathrm{offset}}$}\\
    \bottomrule
  \end{tabular}
\end{table}

\subsection{Transformation}
\label{sec:line-transf}

The simulations show a reasonably well behaved transformation between the true
and recovered slope.  We therefore model the mock data with an affine
transformation, to be able to transform between the measured and true
parameters.

We try to find the best transformation matrix $A$ and vector $\mathbf{b}$
between the measured and true parameters.  For each set of input
($\mathbf{x}_{\mathrm{true}, k}$) and output ($\mathbf{x}_{\mathrm{out}, k}$)
parameters we have:
\begin{align}
  \mathbf{x}_{\mathrm{out}, k} &\approx A \mathbf{x}_{\mathrm{true}, k} + \mathbf{b}
\end{align}

We minimise the function
\begin{align}
  S(A, \mathbf{b}) &= \sum_{k=1}^{N} ||\mathbf{x}_{\mathrm{out}, k} - A \mathbf{x}_{\mathrm{true}, k} - \mathbf{b}||_2^2
\end{align}
with respect to each component of $A$ and $\mathbf{b}$ in order to find the
best-fit transformation $A$ and $\mathbf{b}$ \citep{Spath2004}.  We note that we
do not take the errors on each point $\mathbf{x}_{\mathrm{out}}, k$ into account
as their magnitudes are all comparable (essentially adding a constant to the
equation).

With the best-fit $A$ and $\mathbf{b}$ in hand, we can then invert the equation to
obtain the relation between the observed and the recovered `true' parameters,
which denote as $\mathbf{x}'_{\mathrm{true}}$:
\begin{align}
  \label{eq:inv-trans}
  \mathbf{x}'_{\mathrm{true}} &\approx A^{-1} \left(\mathbf{x}_{\mathrm{out}} - \mathbf{b}\right)\\
  \begin{pmatrix}
    a' \\ c' \\ b' \\ \sigma'_{\mathrm{intr}}
  \end{pmatrix}%
  &= \begin{pmatrix}
      \Amatrixinv
    \end{pmatrix}
    \left(\begin{pmatrix}
      a \\ c \\ b \\ \sigma_{\mathrm{intr}}
    \end{pmatrix}%
  - \begin{pmatrix}
    \tvector
  \end{pmatrix}\right)
\end{align}

\begin{figure*}
  \centering
  \includegraphics[width=0.7\textwidth]{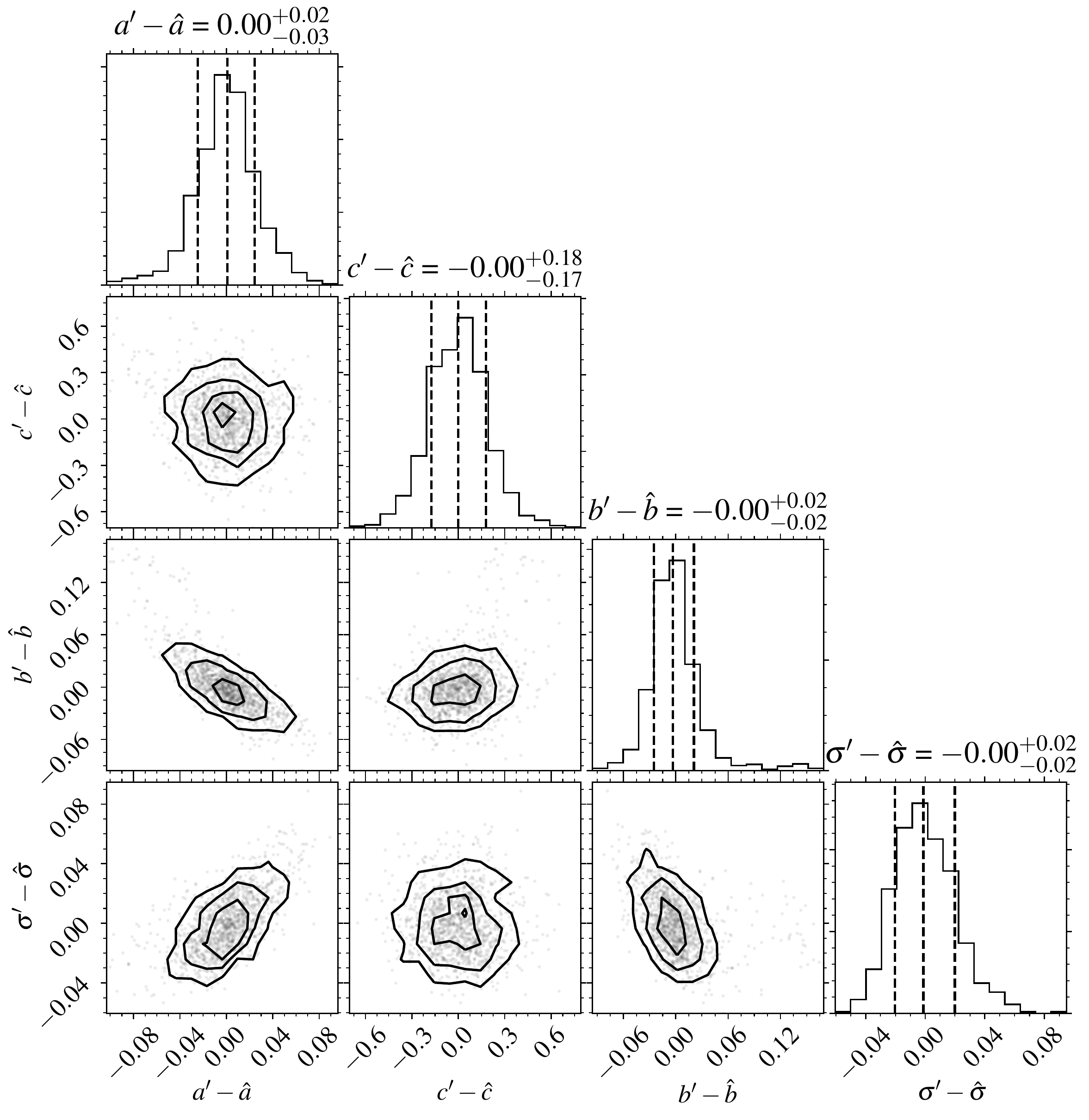}
  \caption{\label{fig:corner-iosim} Plot of the differences between the
    recovered parameters,
    $\mathbf{x}'_{\mathrm{true}} = \left(a', c', b',
      \sigma'_{\mathrm{intr}}\right)^{T}$ and the true parameters,
    $\mathbf{x}_{\mathrm{true}} = \left(\hat a, \hat c, \hat b, \hat
      \sigma_{\mathrm{intr}}\right)^{T}$, for the $N=\simgridsize$ points from
    our simulation; see \Eq{eq:inv-trans}.  We can recover the input parameters
    of our simulation very well, with no mean offset and very small scatter
    (compared to the uncertainty on each parameter obtained from the
    observations). Figure created using the \textsf{corner.py} module
    \protect\citep{Foreman-Mackey2016}.}
\end{figure*}

For our simulated data, we show the distribution of the difference between the
recovered parameters ($\mathbf{x}'_{\mathrm{true}}$) and the true parameters
($\mathbf{x}_{\mathrm{true}}$) in \Fig{fig:corner-iosim}.  We recover the input
parameters very well, with no mean offset between the recovered and the true
parameter.  This shows that the transformation (i.e. $A$ and $\mathbf{b}$) are
very well determined.  Furthermore, the scatter in the differences is much
smaller than the average uncertainty on each parameter obtained from the
observations (of order $\sim 1\%$).  As an illustration, we show the inverse
transformation applied to the simulation by the red lines in \Fig{fig:sfc},
which are now in good agreement with the true values (dashed lines).

In summary, the transformation obtained from the best-fit $A$ and $\mathbf{b}$
is a very accurate description of the bias induced by the flux limit in our
simulated data.  We use the inverse of this transformation, \Eq{eq:inv-trans},
in \Sec{sec:results} to correct our inferred posterior density distribution from
modelling the MUSE data.

\end{appendix}

\end{document}